\documentstyle[preprint,aps,prd]{revtex}
\tightenlines
\begin{document}
\draft
\title{Exclusive Nonleptonic Decays of Bottom and
Charm Baryons in a Relativistic Three-Quark Model:
Evaluation of Nonfactorizing Diagrams}

\author{M.\ A.\ Ivanov}
\address{Bogoliubov Laboratory of Theoretical Physics, Joint Institute
for Nuclear Research, 141980 Dubna (Moscow Region), Russia}

\author{J.\ G.\ K\"{o}rner}
\address{Johannes Gutenberg-Universit\"{a}t, Institut f\"{u}r Physik,
D-55099 Mainz, Germany}

\author{V.\ E.\ Lyubovitskij}
\address{Bogoliubov Laboratory of Theoretical Physics, Joint Institute
for Nuclear Research, 141980 Dubna (Moscow Region), Russia
and Department of Physics, Tomsk State University,
634050 Tomsk, Russia}

\author{A.\ G.\ Rusetsky}
\address{Bogoliubov Laboratory of Theoretical Physics, Joint Institute
for Nuclear Research, 141980 Dubna (Moscow Region), Russia
and IHEP, Tbilisi State University, 380086 Tbilisi, Georgia}

\date{\today}

\maketitle

\begin{abstract}\widetext
Exclusive nonleptonic decays of bottom and charm baryons are studied
within a relativistic three-quark model with a Gaussian shape for the
momentum dependence of the baryon-three-quark vertex. We include
factorizing as well as nonfactorizing contributions to the decay
amplitudes. For heavy-to-light transitions $Q\to q u d$ the total
contribution of the nonfactorizing diagrams amount up to $\sim 60~\%$
of the factorizing contributions in amplitude, and up to $\sim 30~\%$
for $b\to c \bar u d$ transitions. We calculate the rates and the
polarization asymmetry parameters for various nonleptonic decays and
compare them to existing data and to the results of other model
calculations.
\end{abstract}


\section{Introduction}
During the last years there has been significant progress in the
experimental study of nonleptonic decays of heavy baryons~\cite{PDG}.
New results on  the mass spectrum, lifetimes, branching ratios and
asymmetry parameters in the decays of the heavy baryons $\Lambda_c^+$,
$\Sigma_c$, $\Xi_c$, $\Lambda_b^0$, ... were reported by various
experiments ALEPH, ARGUS, ACCMOR, CLEO, OPAL, etc. The heavy baryon
mass spectrum has been determined with good precision (within an accuracy
of a few per cent). As to nonleptonic branching ratios, the accuracy
of the measurements does not exceed 25-30~$\%$ even for the better studied
Cabibbo-favored decay modes $\Lambda^+_c\to\Lambda^0+\pi^+$ and
$\Lambda^+_c\to p+\bar K^0$. For the decay $\Lambda_b^0\to J/\psi \Lambda$
and the Cabibbo-suppressed decay $\Lambda_c^+\to p\phi$ the experimental
errors are even larger. The first observation of the $\Lambda_c^+\to p\phi$
decay was reported by the NA32 Collaboration~\cite{NA32}. They quoted a
branching ratio of
$Br(\Lambda_c^+\to p\phi)/Br(\Lambda_c^+\to pK^-\pi^+) = 0.040 \pm 0.027$.
A more recent measurement of the $\Lambda_c^+\to p\phi$ decay rate by the
CLEO Collaboration resulted in a ratio of branching ratios
$Br(\Lambda_c^+\to p\phi)/Br(\Lambda_c^+\to pK^-\pi^+)=
0.024\pm 0.006\pm 0.003$~\cite{CLEO}. The baryonic decay
$\Lambda_b^0\to J/\psi \Lambda$ was first observed by the UA1
Collaboration~\cite{UA1}. The measured branching ratio was found to be
$Br(\Lambda_b^0\to J/\psi \Lambda) = (1.4 \pm 0.9)~\%$ \cite{PDG}. The OPAL
Collaboration obtained an upper limit for the branching ratio of
$Br(\Lambda_b^0\to J/\psi \Lambda) < 1.1 \%$ \cite{OPAL}. Recently
the CDF Collaboration has reported a much smaller value for the same
quantity $Br(\Lambda_b^0\to J/\psi \Lambda)=(0.037\pm 0.017\pm 0.004)~\%$
from a larger data sample~\cite{CDF}. From a theoretical point of view the
$\Lambda_c^+\to p\phi$ and $\Lambda_b^0\to J/\psi \Lambda$ decays are
simple in as much as they are described by factorizing quark diagrams alone.
Their study can shed light on the nature of the nonleptonic interactions
and may serve as an additional source for determining the
Cabibbo-Kobayashi-Maskawa (CKM) elements and the values of the
short-distance Wilson coefficients in the effective nonleptonic
Lagrangian~\cite{Altarelli}-\cite{Neubert}. In the near future one can
expect large quantities of new data on exclusive charm and bottom baryon
nonleptonic decays which calls for a comprehensive theoretical analysis
of these decays.

There exist a number of theoretical analysis of exclusive nonleptonic heavy
baryon decays in the literature (see, e.g. refs.~\cite{Koerner1}-\cite{Datta})
including predictions for their angular decay distributions.
The analysis of nonleptonic baryon decays is complicated by the necessity
of having to include nonfactorizing contributions. One thus has to go
beyond the factorization approximation which had proved quite useful in
the analysis of the exclusive nonleptonic decays of heavy mesons. There
have been some theoretical attempts to analyse nonleptonic heavy baryon
decays using factorizing contributions alone~\cite{Cheng3}, the argument
being that W-exchange contributions can be neglected in analogy to the
power suppressed W-exchange contributions in the inclusive nonleptonic
decays of heavy baryons. One might even be tempted to drop the
nonfactorizing contributions on account of the fact that they are
superficially proportional to $1/N_c$. However, since $N_c$-baryons
contain $N_c$ quarks an extra combinatorial factor proportional to $N_c$
appears in the amplitudes which cancels the explicit diagrammatic $1/N_c$
factor~\cite{Kramer,SantaFe}. There is now ample empirical evidences in
the $c\to s$ sectors that nonfactorizing diagrams cannot be neglected.
For example, in the charm sector the two observed decays
$\Lambda_c^+\to \Xi^0 K^+$ and $\Lambda_c^+ \to \Sigma \pi$ can only
proceed via nonfactorizing diagrams. Their sizeable observed branching
ratios may thus serve to obtain a measure of the size of the nonfactorizing
contributions.

In the present paper both factorizing and nonfactorizing contributions
to exclusive nonleptonic decays of bottom and charm baryons are taken
into account. The decay amplitudes are studied within a relativistic
three-quark model with a Gaussian shape for the momentum dependence of
the baryon-three-quark vertex. It is shown that the total contribution
of the nonfactorizing diagrams can amount up to $\sim 60~\%$ of the
factorizing contribution for heavy-to-light transitions and up to
$\sim 30~\%$ for $b\to c$ transition in amplitude. We calculate branching
ratios and asymmetry parameters for bottom and charm baryon nonleptonic
decays within the {\it Lagrangian Spectator Model} approach which
generalizes the spectator quark model approach~\cite{Annals,Nucl}.
We compare our results with existing data and other theoretical approaches.

The layout of the paper is as follows. In Section 2 we present details of
our {\it Lagrangian Spectator Model} approach. In Section 3 we discuss the
calculation of the matrix elements of nonleptonic decays of bottom and charm
baryons. In Section 4 we present the results of our calculations.
Section 5 contains our conclusions.

\section{Model}
\baselineskip 20pt

A systematic and comprehensive analysis of weak semileptonic and
nonleptonic decays of heavy baryons has been carried out within the
spectator quark model~\cite{Kramer,DESY,Annals,Nucl} which is based
on the "equal-velocity" approximation~\cite{Annals,Nucl}. Namely,
it is assumed that all quarks inside a hadron have equal velocities
coinciding with the velocity of the hadron. In other words, the internal
relative motion of quarks inside the hadrons is neglected.

The quark-hadron Bethe-Salpeter (BS) wave function satisfies the
free-quark Dirac equation in the each quark index, i.e. the quarks
are assumed to be noninteracting. With the use of the "equal-velocity"
assumption the equations of motion for the wave functions of the
individual constituent quarks in the baryon can be rewritten in terms
of the hadron velocity, thus imposing restriction on the possible form
of the hadronic BS wave function~\cite{Kramer,DESY,Annals,Nucl}. The
explicit form of the BS wave functions for hadron in the initial state
is given by
\begin{eqnarray}\label{BSF}
J^P={\frac{1}{2}}^+&:&\hspace*{2cm}{\cal B}_{ABC} =
\frac{1}{M}\{[(\not\! P + M)\gamma_5 C]_{\beta\gamma}u_\alpha(P)
{\cal B}_{a[bc]}
+ cycl. (\alpha, a; \beta, b; \gamma, c)\},
\nonumber\\
J^P={\frac{3}{2}}^+&:&\hspace*{2cm}{\cal B}_{ABC} =
\frac{1}{M}\{[(\not\! P + M)\gamma_\nu C]_{\beta\gamma}
u^\nu_\alpha(P){\cal B}_{\{abc\}}
+ cycl. (\alpha, a; \beta, b; \gamma, c)\},
\nonumber\\
J^P= 0^{-+}&:&\hspace*{2cm}{\cal M}_A^B =
[(\not\! P + M)\gamma_5]^\beta_\alpha {\cal M}^b_a
\nonumber\\
J^P= 1^{--}&:&\hspace*{2cm}{\cal M}_A^B =
[(\not\! P + M)\not\! \varepsilon]^\beta_\alpha {\cal M}^b_a
\end{eqnarray}
We have suppressed colour indices in (\ref{BSF}). $M$ is the mass of hadron,
$P$ is its total four-momentum, and
${\cal B}_{a[bc]}$, ${\cal B}_{\{abc\}}$,
${\cal M}^b_a$ denote the flavour part of the hadronic wave function.
Analogous formulae for the final state hadronic wave functions
can easily be derived from  Eq.~(\ref{BSF}). They can be
found in ref.~\cite{Kramer}. Note that the BS spin wave functions (1)
contain an additional "projector" factor
\begin{eqnarray}
V_+ = \frac{1}{2M} (\not\! P + M) = \frac{\not\! v + 1}{2}
\end{eqnarray}
\noindent where $v$ is the "on-shell" four-velocity of hadron, i.e.
$v^2 = 1$. The factor $V_+$ ensures that in the c.m. frame only the
positive-energy components of the full BS wave function survive, as it
should indeed be, when the quarks are noninteracting.

Once the explicit form of the hadron wave functions is given, the
transition matrix elements for weak decays are parameterized by a few
overlap integrals in terms of the spin-independent spatial part of the
hadron wave functions. Previously, the overlap integrals have been treated
as phenomenological parameters to be determined from a fit to experimental
data~\cite{Kramer}.

In order to go beyond the approach~\cite{Kramer} one has to develop
a microscopic approach to the overlap integrals appearing in the
expressions for the decay amplitudes or, equivalently, one has to specify
the form of the hadron-quark transition vertex (hadronic BS wave function)
including the explicit momentum dependence of the Lorentz scalar part of
this vertex. In the  Lagrangian model considered in this paper this
dependence is given by the baryon form factor which appears in the nonlocal
interaction vertex coupling the baryons to the three quarks. The Lagrangian
model has been successfully applied to the description of a wide class of the
low and intermediate energy hadron phenomena both in the
light~\cite{Aniv1}-\cite{PSI} and heavy~\cite{Mainz} quark sectors.

In its present form, this model is not immediately applicable to the
study of the heavy baryon nonleptonic decays since it does not reproduce
the results of the spectator model analysis~\cite{Kramer}. The purpose of
our present investigation will consist in embedding, step by step, the
spectator model spin structure in our Lagrangian approach. Put differently,
we attempt to reformulate the spectator model using the Lagrangian language
in order to be able to calculate all quantities appearing in the description
of the nonleptonic decays of heavy baryons with the use of the Feynman
diagram technique.

Let us begin with the formulation of the basic notions of the Lagrangian
model taking into account at every step the spin structure imposed by the
spectator picture.

The problem of the choice of baryonic currents was discussed in
ref.~\cite{Mainz} (see also refs.~\cite{Ioffe}-\cite{QCM}
and~\cite{Shuryak}-\cite{Groote}). Let us briefly review the basic notions.
Suppose that a baryon is a bound state of three quarks. Let  $y_i$ (i=1,2,3)
be the position space four-coordinate of quark $i$ with mass $m_i$.
They are expressed through the center of mass coordinate $(x)$ and the
relative Jacobi coordinates $(\xi_1, ...)$ as
\begin{eqnarray}\label{cmf}
& &y_1=x-3\xi_1\,\,\frac{m_2+m_3}{\sum\limits_{i}m_i}
\nonumber\\
& &y_2=x+3\xi_1\,\,\frac{m_1}{\sum\limits_{i}m_i}-
2\xi_2\sqrt{3}\,\,\frac{m_3}{m_2+m_3}
\\
& &y_3=x+3\xi_1\,\,\frac{m_1}{\sum\limits_{i}m_i}+
2\xi_2\sqrt{3}\,\,\frac{m_2}{m_2+m_3}
\nonumber\\
& &\nonumber\\
\text{where\hspace*{1cm}}
& &x=\frac{\sum\limits_{i}m_iy_i}{\sum\limits_{i}m_i},
\hspace{0.5cm}
\xi_1=\frac{1}{3}\,\,\bigg(\frac{m_2y_2+m_3y_3}{m_2+m_3}-y_1\biggr),
\hspace{0.5cm}
\xi_2=\frac{y_3-y_2}{2\sqrt{3}}.
\nonumber
\end{eqnarray}

In the case of light baryons we shall work in the limit of SU(3) invariance
by assuming that the masses of $u, d$ and $s$ quarks are equal to each other
in Eq.~(\ref{cmf}). The breaking of SU(3) symmetry through the position space
variables $y_i$ (via a difference of strange $m_s$ and nonstrange $m$ quark
masses: $m_s-m\neq 0$) was found to be insignificant~\cite{Mainz}.
Thus, for light baryons composed of $u,d$ or $s$ quarks
the coordinates of the quarks may be written as
\begin{eqnarray}
& &y_1=x-2\xi_1 \hspace{0.5cm} y_2=x+\xi_1-\xi_2\sqrt{3}
\hspace{0.5cm} y_3=x+\xi_1+\xi_2\sqrt{3} \nonumber
\end{eqnarray}

For a heavy-light baryon with $m_1\gg m_2,\,m_3$ one has instead
\begin{eqnarray}
y_1=y_Q=x, \hspace*{1cm} y_2=y_{q_1}=x+3\xi_1-\xi_2\sqrt{3},
           \hspace*{1cm} y_3=y_{q_2}=x+3\xi_1+\xi_2\sqrt{3}
\nonumber
\end{eqnarray}

We assume that the momentum distribution of the constituents inside
a baryon is modelled by an effective relativistic vertex function given
by $$F\left(\frac{\Lambda_B^2}{18}\sum\limits_{i<j}(y_i-y_j)^2\right)$$
which depends only on the sum of the relative coordinates squared in
the coordinate space and on a cutoff parameter $\Lambda_B$. Generally
speaking, the shape of this function should be determined from the
bound state equation and may depend on the flavours of the quarks
involved. In order to reduce the number of free parameters we will
use a common Gaussian function for all flavours but we allow for flavour
dependent values of the cutoff parameter $\Lambda_B$. The Gaussian
shape guarantees ultraviolet convergence of the matrix elements.
The vertex function models the long distance QCD interactions between
quarks. For the present application, there are at least three
different values for $\Lambda_B$ corresponding to the $(s,d,u)$, $(c,d,u)$,
and $(b,d,u)$ sectors. However, in order to recover the Isgur-Wise
symmetry  in the heavy quark limit ($m_Q\to\infty$) the cutoff parameter
$\Lambda_B$ has to be the same for charm and bottom baryons.

The Lagrangian describing the interaction of baryons with
the three-quark current is written as
\begin{eqnarray}\label{strong}
{\cal L}_B^{\rm int}(x)&=&g_B\bar B(x)
\hspace*{-.1cm}\int \hspace*{-.1cm}dy_1\hspace*{-.1cm}\int
\hspace*{-.1cm}dy_2\hspace*{-.1cm}\int \hspace*{-.1cm}dy_3 \,
\delta\left(x-\frac{\sum\limits_i m_iy_i}{\sum\limits_i m_i}\right)
F\left(\frac{\Lambda_B^2}{18}\sum\limits_{i<j}(y_i-y_j)^2\right)\nonumber \\
& &\nonumber\\
&\times&J_B(y_1,y_2,y_3)+h.c.
\end{eqnarray}
where $J_B(y_1,y_2,y_3)$ is the three-quark current with quantum numbers
of a baryon $B$:
\begin{equation}\label{current}
J_B(y_1,y_2,y_3)=\Gamma_1 q^{a_1}(y_1)q^{a_2}(y_2)C\Gamma_2 q^{a_3}(y_3)
\varepsilon^{a_1a_2a_3}.
\end{equation}
\noindent Here $\Gamma_{1,2}$ are strings of Dirac matrices,
$C=\gamma^0\gamma^2$ is the charge conjugation matrix and $a_i$
are the color indices. The strong coupling constant $g_B$
in (\ref{strong}) can be calculated from {\it the compositeness condition}
(see, ref.~\cite{Mainz}, \cite{Lubovit}-\cite{QCM}), i.e.
the renormalization constant of the hadron
wave function is set equal to zero, $Z_H=1-g_H^2\Sigma^\prime_B(M_H)=0$,
with $\Sigma_H$ being the hadron mass operator and $M_H$ denotes a hadron
mass. Note that the latter condition is equivalent to the well-known
relativistic normalization condition for the hadronic Bethe-Salpeter (BS)
wave function. However, for technical reasons it is more convenient to use
the normalization condition for the elastic vector form factor at zero
recoil which, of course, is completely equivalent to the compositeness
condition (see, discussion about it in ref.~\cite{Mainz}).

Possible choices of light and heavy-light baryonic
currents have been studied in refs.~\cite{Ioffe}-\cite{QCM}
and~\cite{Shuryak}-\cite{Groote}. For the octet of light baryons, for
the $\Lambda$-type heavy-light baryons ($\Lambda_Q$, $\Xi_Q$) with a light
spin zero diquark system, and for the $\Omega$-type heavy-light baryons
($\Omega_Q$, $\Sigma_Q$) with a light spin one diquark system the
currents are written as follows~\cite{Mainz}.

\begin{eqnarray}
\begin{array}{ll}
\text{Light Baryon Currents} & \\
\text{\it {vector variant}} &
\hspace*{-1cm}
J_B^V(y_1,y_2,y_3)=\gamma^\mu\gamma^5 q^{a_1}(y_1)q^{a_2}(y_2)
C\gamma_\mu q^{a_3}(y_3)\varepsilon^{a_1a_2a_3}
\label{currents_l}\\
\text{\it {tensor variant}} &
\hspace*{-1cm}
J_B^T(y_1,y_2,y_3)=\sigma^{\mu\nu}\gamma^5 q^{a_1}(y_1)q^{a_2}(y_2)
C\sigma_{\mu\nu} q^{a_3}(y_3)\varepsilon^{a_1a_2a_3}\\
\end{array}
\end{eqnarray}
\begin{eqnarray}
\begin{array}{ll}
\text{Heavy-Light Baryon Currents} & \\
\text{{\it pseudoscalar variant}} &
\hspace*{-1cm}
J_{\Lambda_Q}^P=\varepsilon^{abc}Q^au^bC\gamma^5d^c\\
\text{{\it axial variant}} &
\hspace*{-1cm}
J_{\Lambda_Q}^A=\varepsilon^{abc}\gamma_\mu Q^au^b
C\gamma^\mu\gamma^5d^c\label{currents_h}\\
\text{{\it vector variant}} &
\hspace*{-1cm}
J_{\Omega_Q}^{V}=\varepsilon^{abc}\gamma_\mu\gamma^5Q^a
s^bC\gamma^\mu s^c,\,\,\, J_{\Omega^\star_Q}^{V;\mu}=\varepsilon^{abc}Q^a
s^bC\gamma^\mu s^c\\
\text{{\it tensor variant}} &
\hspace*{-1cm}
J_{\Omega_Q}^{T}=\varepsilon^{abc}\sigma_{\mu\nu}\gamma_5Q^a
s^bC\sigma^{\mu\nu}s^c,\,\,\, J_{\Omega^\star_Q}^{T;\mu}=
-i\varepsilon^{abc}\gamma_\nu Q^as^bC\sigma^{\mu\nu} s^c
\end{array}
\end{eqnarray}
\noindent
In Table I we give the quark content, the quantum numbers (spin-parity $J^P$,
spin $S_{qq}$ and isospin $I_{qq}$ of light diquark) and the experimental
(when available) and theoretical mass spectrum of the heavy baryons
\cite{PDG,DESY} analyzed in this paper. Square brackets $[...]$ and curly
brackets $\{...\}$ denote antisymmetric and symmetric flavour and spin
combinations of the light degrees of freedom. The masses of the light baryons
are taken from the Review of Particle Properties \cite{PDG}.

Next we write down the Lagrangian which describes the interaction of
$\Lambda_Q$-baryon with quarks in the heavy quark limit ($m_Q\to\infty$),
i.e. to leading order in the $1/m_Q$ expansion
\begin{eqnarray}\label{lagr_LB}
{\cal L}_{\Lambda_Q}^{\rm int}(x)&=&g_{\Lambda_Q}\bar \Lambda_Q(x)
\Gamma_1 Q^a(x)\int d\xi_1 \int d\xi_2\,\,\,
F(\Lambda_{B_Q}^2\cdot[\xi_1^2+\xi_2^2])\\
&\times&u^b(x+3\xi_1-\xi_2\sqrt{3})C\Gamma_2d^c(x+3\xi_1+\xi_2\sqrt{3})
\varepsilon^{abc}+h.c.
\nonumber
\end{eqnarray}
\noindent where
\[\Gamma_1\otimes C\Gamma_2=\left\{
\begin{array}{ll}
I\otimes C\gamma^5
&\,\,\,\,\text{pseudoscalar current}\\[3mm]
\gamma_\mu \otimes C\gamma^\mu\gamma^5
&\,\,\,\,\text{axial current}
\end{array}\right. \]
One can see that the heavy quark is factorized from the light degrees of
freedom in this limit. The vertex form factor $F$ characterizes the
distribution of $u$ and $d$ quarks inside the $\Lambda_Q$ baryon.
It is readily seen that the Lagrangian (\ref{lagr_LB}) exhibits the
heavy quark flavour symmetry (symmetry under exchange b with c)
if the parameter $\Lambda_{B_Q}$ is the same for charm and bottom baryons.

In what follows we shall work with the momentum space representation of the
interaction Lagrangians. Performing the requisite Fourier transformation
e.g. for the case of the $\Lambda_Q$ baryon we obtain
\begin{eqnarray}\label{mom_LB}
{\cal L}_{\Lambda_Q}^{\rm int}(p)&=&g_{\Lambda_{B_Q}}
\bar \Lambda_Q(p)
\int\hspace*{-.1cm}dp_1\int\hspace*{-.1cm}dp_2\int\hspace*{-.1cm}dp_3
\int\hspace*{-.1cm}dk_1\int\hspace*{-.1cm}dk_2\,
\delta(k_1-3(p_2+p_3))\,\,\delta(k_2-\sqrt{3}(p_3-p_2))
\nonumber\\
& &\nonumber\\
&\times&
\delta(p-\sum\limits_ip_i)
F\biggl(\frac{k_1^2+k_2^2}{\Lambda^2_{B_Q}}\biggr)
\Gamma_1 Q^a(p_1)u^b(p_2)C\Gamma_2d^c(p_3)\varepsilon^{abc}+h.c.
\end{eqnarray}
\noindent
where $p$ and $p_1, p_2, p_3$ are the momenta of the baryon and the
constituent quarks, respectively. The relative momenta $k_1$ and $k_2$
may be expressed in terms of the quark momenta $p_i$ in a standard
manner~\cite{Mainz}.

For our purposes we also need the effective Lagrangians that describe the
coupling of pions, kaons and the vector mesons $\rho$, $\phi$ and $J/\psi$
to their quark constituents. In this paper we also assume that the mesons
are point-like objects, i.e. their interaction with the constituent quarks
are described by a local nonderivative Lagrangian
\begin{eqnarray}\label{Meson}
L_M (p) = g_M M(p) \int dp_1 \int dp_2 \,\, \delta(p-p_1-p_2) \,\,
\bar q(p_1) \,\, \Gamma_M \,\, \lambda_M \,\, q(p_2) + h.c.
\end{eqnarray}
\noindent where $\Gamma_M$ and $\lambda_M$ are spin and flavour matrices.
In other words, we choose the effective meson vertex functions to be
constants in momentum space. This is a reliable approximation for the light
mesons. For heavy mesons we expect that form factor effects in the meson
vertex become important. This prevents us from extending the present approach
to cases with heavy mesons in the final states, such as
$\Lambda_b^0\rightarrow\Lambda_c^++D_s^-$. In general
the form factor effects in the decays involving heavy mesons in the final
state are expected to suppress their rates relative to those obtained from
a point-like vertex. Exclusive nonleptonic bottom baryon decays involving
heavy mesons form the subject of a separate piece of work.

To reproduce the spin amplitude structure of the spectator (or static quark)
model analysis~\cite{Kramer,DESY} we assign the projector
$V_+ = (\not\!\! v + 1) / 2$ to each light quark field in the baryon-quark
vertex, where $v$ is the "on-shell" four-velocity of hadron as in
ref.~\cite{Kramer}. The conjugate antiquark fields in the mesons are
multiplied by the projector $V_- = (- \not\! v + 1)/2$.
We shall also use the static approximation for
$u$, $d$ and $s$ quark propagators
\begin{eqnarray}\label{static_prop}
<0|T\{q(x)\bar q(y)\}|0> = \frac{1}{\Lambda_q} \,\, \delta^{(4)}(x-y)
\end{eqnarray}
\noindent where $\Lambda_q$ is the free parameter having the dimension
of mass. We choose this parameter to have the same value $\Lambda$ for
$u$ and $d$ quarks and a different value $\Lambda_s$ for the strange quark.
The model obtained with the use of above prescriptions will be referred
to as {\it Lagrangian Spectator Model} in what follows.

An important property of the Lagrangian Spectator Model is that the
structure of the interaction Lagrangians of light and heavy-light
baryons with quarks is simplified. Namely, the different options for the
choice of baryon currents all become equivalent. For example, the vector
and tensor forms of the interaction Lagrangians of $J^P=1/2^+$ light baryons
are completely equivalent. For the proton the interaction Lagrangian takes
the form
\begin{eqnarray}\label{lagr-pr}
{\cal L}_P^{\rm int}(p)
&=&4g_p\bar p(p)\hspace*{-.1cm}\int \hspace*{-.1cm}dk_1\hspace*{-.1cm}
\int \hspace*{-.1cm}dk_2
F\biggl(12\frac{k_1^2+k_2^2+k_1k_2}{\Lambda_{B_q}^2}\biggr)\\
& &\nonumber\\
&\times&V_+u^{a_1}(k_1+p)u^{a_2}(k_2)C\gamma_5V_+d^{a_3}(-k_1-k_2)
\varepsilon^{a_1a_2a_3}+h.c.\nonumber\\
& &\nonumber\\
&\equiv&2g_p\bar p(p)\hspace*{-.1cm}\int \hspace*{-.1cm}dk_1
\hspace*{-.1cm}\int\hspace*{-.1cm}dk_2
F\biggl(12\frac{k_1^2+k_2^2+k_1k_2}{\Lambda_{B_q}^2}\biggr)\nonumber\\
& &\nonumber\\
&\times&V_+\gamma^\mu\gamma^5d^{a_1}(k_1+p)u^{a_2}(k_2)C\gamma_\mu V_+
u^{a_3}(-k_1-k_2)\varepsilon^{a_1a_2a_3}+h.c.
\nonumber
\end{eqnarray}
\noindent
In Appendix A we provide a full list of the effective interaction
Lagrangians for light baryons in the Lagrangian Spectator Model.

In the Lagrangian Spectator Model the leptonic coupling
constants $f_\pi$ and $f_K$ are determined by the integrals
\begin{eqnarray}\label{fpifk}
f_\pi = \frac{N_cg_\pi}{4\pi^2} \frac{1}{M_\pi \Lambda^2}
\int_{reg} \frac{d^4 k}{\pi^2}, \hspace*{2cm}
f_K   = \frac{N_cg_K}{4\pi^2} \frac{1}{M_K \Lambda\Lambda_s}
\int_{reg} \frac{d^4 k}{\pi^2}\label{fpik}
\end{eqnarray}
\noindent The meson coupling constants $g_\pi$ and $g_K$ in
Eq.~\ref{fpifk}) are determined from
{\it the compositeness condition}~\cite{Mainz} which reads
\begin{eqnarray}
1 = \frac{N_cg_\pi^2}{4\pi^2} \frac{1}{M_\pi^2 \Lambda^2}
\int_{reg} \frac{d^4 k}{\pi^2}, \hspace*{2cm}
1 = \frac{N_cg_K^2}{4\pi^2} \frac{1}{M_K^2 \Lambda\Lambda_s}
\int_{reg} \frac{d^4 k}{\pi^2}\label{gpik}
\end{eqnarray}
\noindent Equations (\ref{fpik}) and (\ref{gpik}) contain the ultraviolet
divergence since the mesons in our scheme are point-like objects.
To regularize these quantities we introduce an ultraviolet cutoff parameter
$\Lambda_{cut}$. In order to reduce the number of free parameters in the
model we relate the cutoff parameter in Eqs.~(\ref{fpifk}) and (\ref{gpik})
to the parameters $\Lambda$ and $\Lambda_s$ appearing in static light quark
propagator (\ref{static_prop}) via
$\Lambda_{cut}=\Lambda_{q_1}\Lambda_{q_2}/(\Lambda_{q_1}+\Lambda_{q_2})$.
Here $q_i$ corresponds to the flavour of the light quark being
the constituent. After that we get
\begin{eqnarray}\label{lepcon}
f_\pi = \frac{\sqrt{N_c}}{8\pi} \Lambda, \hspace*{2cm}
f_K = \frac{\sqrt{N_c}}{2\pi} \frac{(\Lambda\Lambda_s)^{3/2}}
{(\Lambda+\Lambda_s)^2}
\end{eqnarray}
\noindent Substituting experimental values for $f_\pi$ = 131 MeV
and $f_K$ = 160 MeV in Eqs.~(\ref{lepcon}) we obtain $\Lambda$=1.90 GeV
and $\Lambda_s$=3.29 GeV.

For the heavy quark propagator $S_Q$ we will use the leading term in the
inverse mass expansion. Suppose $p=M_{B_Q}v$ is the heavy baryon momentum.
We introduce the parameter $\bar\Lambda_{\{q_1q_2\}}=M_{\{Qq_1q_2\}}-m_Q$
which is the difference between the heavy baryon mass
$M_{\{Qq_1q_2\}}\equiv M_{B_Q}$
and the heavy quark mass. Keeping in mind that the vertex function
falls off sufficiently fast such that the condition
$|k|<<m_Q$ holds ($k$ is the virtual momentum of light quarks) one has
\begin{eqnarray}\label{Sheavy}
& &S_Q(p+k)={1\over m_Q - (\not\! p \,\, + \not\! k)}=
\frac {m_Q+M_{B_Q}\not\! v+\not\! k} {m_Q^2-M_{B_Q}^2-2M_{B_Q}vk-k^2}
=S_v(k,\bar\Lambda_{\{q_1q_2\}})+O\biggl(\frac{1}{m_Q}\biggr)
\nonumber\\[5mm]
& &S_v(k,\bar\Lambda_{\{q_1q_2\}}) = - \frac{(1+\not\! v)}
{2(v\cdot k + \bar\Lambda_{\{q_1q_2\}})}
\end{eqnarray}
In what follows we will assume that
$\bar\Lambda\equiv\bar\Lambda_{uu}=\bar\Lambda_{dd}=\bar\Lambda_{du}$,
$\bar\Lambda_{s}\equiv\bar\Lambda_{us}=\bar\Lambda_{ds}$. Thus there are
altogether three independent parameters: $\bar\Lambda$, $\bar\Lambda_{s}$,
and $\bar\Lambda_{ss}$.

The vertex function $F$ in the baryon-quark interaction Lagrangians is an
arbitrary function except that it should render the Feynman diagrams
ultraviolet finite as was mentioned before. In~\cite{Aniv1}-\cite{Mainz}
it was found that the basic physical observables of pion and nucleon
low-energy physics depend only weakly on the choice of the vertex functions.
In the present paper we choose a Gaussian vertex function for simplicity.
In Minkowski space we write
\begin{eqnarray}
F\biggl(\frac{k^2_1+k^2_2}{\Lambda_B^2}\biggr)&=&
\exp\biggl(\frac{k^2_1+k^2_2}{\Lambda_B^2}\biggr)
\nonumber
\end{eqnarray}
where $\Lambda_B$ is the Gaussian range parameter which is related to the
size of a baryon. Note that all calculations are done in the Euclidean region
($k^2_i=-k^2_{i E}$) where the above vertex function decreases very rapidly.
We consider two different values of the $\Lambda_B$ cutoff parameter:
$\Lambda_{B_q}$ for light baryons composed from light $(u, d, s)$ quarks
and $\Lambda_{B_Q}$ for baryons containing a single heavy quark ($b$ or $c$).
The requirement of the unit normalization of the baryonic IW-functions
$\zeta(\omega)$ and $\xi_1(\omega)$ at zero recoil $\omega=1$ ($\zeta(1)=1$,
$\xi_1(1)=1$) imposes the restriction $\Lambda_{B_b}=\Lambda_{B_c}$. This can
be seen by expressing the baryonic IW-functions for arbitrary values of
$\Lambda_{B_Q}$ as
\begin{eqnarray}\label{zeta_om}
\left.\zeta(\omega)=
\frac{\Phi\biggl(\sqrt{2\Lambda_{B_b}^2\Lambda_{B_c}^2/
(\Lambda_{B_b}^2+\Lambda_{B_c}^2)},\omega\biggr)}
{\sqrt{\Phi(\Lambda_{B_b},1)}\sqrt{\Phi(\Lambda_{B_c},1)}}\right.
\end{eqnarray}
\begin{eqnarray}
& &\Phi(\Lambda_{B_Q},w)=\Lambda_{B_Q}^6
(\omega + 1) \int\limits_0^\infty duu\int\limits_0^1 dx
\exp\biggl[-18u^2 -36u^2x(1-x)(\omega-1)
+ 36u\frac{\bar\Lambda}{\Lambda_{B_Q}}
\biggr]
\nonumber
\end{eqnarray}
\noindent Eq.~(\ref{zeta_om}) shows that one recovers $\zeta(1)=1$ only when
$\Lambda_{B_b} = \Lambda_{B_c}$. As was mentioned above, the parameter
$\Lambda_{B_Q} = \Lambda_{B_b} = \Lambda_{B_c}$ is one of the adjustable
parameters in our calculation.

Thus, there is the following set of adjustable parameters in our model:
the cutoff parameters $\Lambda_B$ ($\Lambda_{B_q}$ and $\Lambda_{B_Q}$),
and a set of $\bar\Lambda_{\{q_1q_2\}}$ binding energy parameters:
$\bar\Lambda$, $\bar\Lambda_s$ and $\bar\Lambda_{\{ss\}}$.

\section{Matrix Elements of Weak Decays of Heavy Baryons}

The weak nonleptonic decays of bottom and charm baryons are described by the
diagrams I, IIa, IIb and III in Fig. 1.
\footnote
{In the terminology of~\cite{Cheng3} diagram I corresponds to factorizable
external and internal W-emission, IIa to nonfactorizable internal W-emission
and IIb and III to nonfactorizable W-exchange.}

Diagram I corresponds to the so-called factorizing contribution. Diagrams
IIa, IIb and III correspond to the nonfactorizing
contributions. The vertices
$O_\mu\bullet \bullet O_\mu$ correspond to the nonleptonic interaction
described by a standard effective four-fermion Lagrangian
\cite{Altarelli}-\cite{Neubert}. For $b\to c \bar u d$ and
$c \to s \bar u d$ transitions the effective four-fermion  vertices read
\footnote
{We employ the notation
\[ \gamma_5=\left(
\begin{array}{cc}
0  & -I \\
-I & 0
\end{array} \right)\]
}
\begin{eqnarray}\label{eff}
{\cal L}_{eff}&=&\frac{G_F}{\sqrt{2}}V_{cb}V^{\dag}_{ud}
[c_1(\bar c^{a_1} O_\mu b^{a_1}) (\bar d^{a_2} O_\mu u^{a_2})
+ c_2(\bar c^{a_1} O_\mu b^{a_2}) (\bar d^{a_2} O_\mu u^{a_1})]
\\
& &\nonumber\\
&+&\frac{G_F}{\sqrt{2}}V_{cs}V^{\dag}_{ud}
[c_1^\star(\bar s^{a_1} O_\mu c^{a_1}) (\bar u^{a_2} O_\mu d^{a_2})
+ c_2^\star(\bar s^{a_1} O_\mu c^{a_2})
(\bar u^{a_2} O_\mu d^{a_1}) ] + h.c.,   \nonumber\\
& &\nonumber\\
O_\mu &=& \gamma_\mu(1+\gamma_5)
\nonumber
\end{eqnarray}
\noindent Here $c_1, c_2$ are short distance Wilson coefficients for
$b\to c\bar u d$ transitions and $c^\star_1, c^\star_2$ are the Wilson
coefficients for $c\to s\bar u d$ transitions. It is well-known that
the factorizing contributions are proportional to the following two
linear combinations
\begin{eqnarray}
a_1&=&c_1 + \frac{c_2}{N_c} = c_1 + \xi c_2\label{a_ch}\\
a_2&=&c_2 + \frac{c_1}{N_c} = c_2 + \xi c_1\label{a_bt}
\end{eqnarray}
\noindent and the same for $a^\star_1$ and $a^\star_2$. Here $N_c$ is
the number of colors and $\xi=1/N_c$ is the color singlet projection
factor. Phenomenological considerations of the nonleptonic
decays of $D$ and $B$ mesons give the following values for the Wilson
coefficients
\begin{eqnarray}
& &a^\star_1 \approx 1.2 \pm 0.10 \approx c_1, \,\,\,\,\,
a^\star_2 \approx -0.5 \pm 0.10 \approx c_2 \,\,\,\text{\cite{Neubert}}
\nonumber\\
& &a_1 \approx 1.05 \pm 0.10, \,\,\,\,\,
a_2 \approx 0.25 \pm 0.05 \,\,\, \text{(see e.g. refs. in \cite{Buras})}
\nonumber
\end{eqnarray}
\noindent
The phenomenological results for the coefficients $a^\star_{1,2}$
can be seen to correspond to a suppression of the $1/N_c$ term in
Eq.~(\ref{a_ch}). A straightforward calculation of these coefficients
in the leading logarithmic approximation has been performed in
refs.~\cite{Altarelli,Gaillard,Buras}. For D-meson decays it was shown
that the coefficient $a^\star_1$ is weakly dependent on
the choice of the renormalization scheme for fixed values of the
renormalization scale and the QCD cutoff parameter:
$a^\star_1$ = 1.31 $\pm$ 0.19 (in accordance with phenomenology). In contrast
to this the value of $a^\star_2$ strongly depends on the renormalization
scheme, ranging from -0.47$\pm$0.15 to -0.60$\pm$0.22. A detailed discussion
can be found in ref.~\cite{Buras}. A first calculation of the Wilson
coefficients $a_i$ for bottom hadron decays was done in
ref.~\cite{Altarelli,Gaillard}. A more refined analysis of the renormalization coefficients within various renormalization
schemes can be found in ref. \cite{Buras} where was shown that the value of
the coefficient $a_1$ depends weakly on details of calculations:
$a_1$=1.01$\pm$0.02 (in accordance with phenomenological analysis). As for
the case of charm decays the coefficient $a_2$ is more sensitive to the
choice of the renormalization scheme and ranges from
0.15$\pm$0.05 to 0.20$\pm$0.05.

The matrix elements describing heavy-to-heavy ($b\to c$) and
heavy-to-light ($Q\to q$) transitions can be written as

\vspace*{1cm}
$\bullet$ heavy-to-heavy transition

\vspace*{1cm}
\underline{Factorizing contribution}

\vspace*{.5cm}
Diagram I
\begin{eqnarray}\label{BCFD-trans}
T^{fac}_{B_b\to B_c + M} &=&
\frac{G_F}{\sqrt{2}} \,\,V_{cb} \,\,V^\dagger_{q_1q_2}\,\,
\chi_{\pm} <B_c|J_\mu^{V+A}|B_b> \cdot <M|J_\mu^{V+A}|0>, \nonumber\\[5mm]
<B_c|J_\mu^{V+A}|B_b> &=& \frac{N_c!g^2_{B_Q}}{(4\pi)^4\Lambda_{q_1}
\Lambda_{q_2}}\int\frac{d^4k}{\pi^2i}\int\frac{d^4k^\prime}{\pi^2i}
\hspace*{0.1cm}
\exp\biggl[\frac{18k^2+6(2k^\prime+k)^2}{\Lambda^2_{B_Q}}\biggr]
\\[5mm]
&\times&\bar u(v_2)\Gamma_2 S_{v_2}(k,\bar\Lambda) O^\mu
S_{v_1}(k,\bar\Lambda)\Gamma_1 u(v_1)\,\,
{\rm Tr}\bigg[\Gamma_2^\prime (1 + \not\! v_2)
(1 + \not\! v_1)\Gamma_1^\prime \biggr]
 \nonumber
\end{eqnarray}
\noindent For the matrix elements of the current operator $J_\mu^{V+A}$
sandwiched between one-meson state $<M|$ and the vacuum $|0>$ we use the
standard definitions
\begin{eqnarray}
<M^P(P_3)|A^\mu|0> &=& f_P P_3^\mu
\hspace*{2.5cm} \text{for the pseudoscalar
mesons}\nonumber\\
<M^V(P_3)|V^\mu|0> &=& f_V M_3 \varepsilon^\mu  \hspace*{2cm}
\text{for the vector mesons}\nonumber
\end{eqnarray}
\noindent Here $\chi_+=a_1$ for transition with a charged meson in the
final state and $\chi_-=a_2$ for transition with a neutral meson in the
final state. $P_3$ and $M_3$ are the four-momentum and the mass of the
meson, respectively, $f_P$ is the leptonic decay constant of pseudoscalar
meson and $f_V$ is the decay constant of vector meson into $e^+ e^-$ pair.
For $f_P$ and $f_V$ we use the experimental values \cite{PDG}:
$f_\pi$ = 131 MeV, $f_K$ = 160 MeV, $f_\phi$ = 237 MeV,
$f_{J/\psi}$ = 405 MeV.

\vspace*{1cm}
\underline{Nonfactorizing contributions}

\vspace*{.5cm}
Diagram IIa
\begin{eqnarray}\label{BCIIa-trans}
T^{IIa}_{B_b\to B_c + M} &=&
\frac{G_F}{\sqrt{2}} \,\, V_{cb} \,\, V^\dagger_{q_1q_2} \,\,
\frac{N_c!\,\,g^2_{B_Q}\,\,g_M}{(4\pi)^6\Lambda_{q_1}\cdots \Lambda_{q_4}}
\int\frac{d^4k}{\pi^2i} \int\frac{d^4k^\prime}{\pi^2i}
\int\frac{d^4k^{\prime\prime}}{\pi^2i}\\[5mm]
&\times&\exp\biggl[\frac{9k^2+9k^{\prime 2}+3(2k^{\prime\prime}+2k^\prime
-k-p_3)^2+3(2k^{\prime\prime}+k^\prime - p_3)^2}{\Lambda^2_{B_Q}}\biggr]
\nonumber\\[5mm]
&\times&\bar u(v_2)\Gamma_2 S_{v_2}(k,\bar\Lambda) O^\mu
S_{v_1}(k,\bar\Lambda)\Gamma_1 u(v_1)\,\,
{\rm Tr}\bigg[\Gamma_2^\prime (1 + \not\! v_2) \Gamma_M
(1 + \not\! v_3) O_\mu (1 + \not\! v_1) \Gamma_1^\prime\biggr]
\nonumber
\end{eqnarray}
\noindent Here $g_M$ is the meson-quark coupling constant which is
calculated with the use of the compositeness condition.
The Dirac structure $\Gamma_M$ specifies the mesonic final state, i.e.
$\Gamma_M=i\gamma_5$ for pseudoscalar mesons and $\Gamma_M=\gamma_\mu$ for
vector mesons

\vspace*{.5cm}
Diagram IIb
\begin{eqnarray}\label{BCIIb-trans}
T^{IIb}_{B_b\to B_c + M} &=&
\frac{G_F}{\sqrt{2}} \,\, V_{cb} \,\, V^\dagger_{q_1q_2} \,\,
\frac{N_c!\,\,g^2_{B_Q}\,\,g_M}{(4\pi)^6\Lambda_{q_1}\cdots \Lambda_{q_4}}
\int\frac{d^4k}{\pi^2i} \int\frac{d^4k^\prime}{\pi^2i}
\int\frac{d^4k^{\prime\prime}}{\pi^2i}\\[5mm]
&\times&\exp\biggl[\frac{9k^2+9k^{\prime 2}+3(2k^{\prime\prime}+k+p_3)^2
+3(2k^{\prime\prime}+2k-k^\prime + p_3)^2}{\Lambda^2_{B_Q}}\biggr]
\nonumber\\[5mm]
&\times&\bar u(v_2)\Gamma_2 S_{v_2}(k,\bar\Lambda) O^\mu
S_{v_1}(k,\bar\Lambda)\Gamma_1 u(v_1)\,\,
{\rm Tr}\bigg[\Gamma_2^\prime (1 + \not\! v_2) O_\mu \Gamma_M
(1 + \not\! v_3) (1 + \not\! v_1) \Gamma_1^\prime\biggr]
\nonumber
\end{eqnarray}

\newpage
Diagram III
\begin{eqnarray}\label{BCIII-trans}
T^{III}_{B_b\to B_c + M} &=& \frac{G_F}{\sqrt{2}} V_{cb} V^\dagger_{q_1q_2}
\frac{N_c!g^2_{B_Q}g_M}{(4\pi)^6\Lambda_{q_1}\cdots \Lambda_{q_4}}
\int\frac{d^4k}{\pi^2i}\int\frac{d^4k^\prime}{\pi^2i}
\int\frac{d^4k^{\prime\prime}}{\pi^2i}\hspace*{0.1cm}\\[5mm]
&\times&\exp\biggl[\frac{9k^2+9k^{\prime 2}+3(2k^{\prime\prime}-k-p_3)^2
+3(2k^{\prime\prime}-k^\prime + p_3)^2}{\Lambda^2_{B_Q}}\biggr]
\nonumber\\[5mm]
&\times&\bar u(v_2)\Gamma_2 S_{v_2}(k,\bar\Lambda) O^\mu
S_{v_1}(k,\bar\Lambda)\Gamma_1 u(v_1)\,\,
{\rm Tr}\bigg[\Gamma_2^\prime (1 + \not\! v_2) O_\mu
(1 + \not\! v_1)\Gamma_1^\prime \Gamma_M(1 + \not\! v_3)\biggr]
\nonumber
\end{eqnarray}

\vspace*{1cm}
$\bullet$ heavy-to-light transition

\vspace*{1cm}
\underline{Factorizing contribution}

\vspace*{.5cm}
Diagram I
\begin{eqnarray}\label{QqFD-trans}
T^{fac}_{B_Q\to B_q + M} &=&
\frac{G_F}{\sqrt{2}} \,\,V_{Qq} \,\,V^\dagger_{q_1q_2}\,\,
\chi_{\pm} <B_q|J_\mu^{V+A}|B_Q> \cdot <M|J_\mu^{V+A}|0>, \nonumber\\[5mm]
<B_q|J_\mu^{V+A}|B_Q> &=& \frac{N_c!\,\,g_{B_Q}\,\,g_{B_q}}
{(4\pi)^4\Lambda_q\Lambda_{q_1}\Lambda_{q_2}}
\int\frac{d^4k}{\pi^2i}\int\frac{d^4k^\prime}{\pi^2i}\hspace*{0.1cm}
\exp\biggl[\frac{9k^2+3(2k^\prime+k)^2}{\Lambda^2_{B_Q}}\biggr]
\\[5mm]&\times&
\exp\biggl[\frac{(3k+2p_2)^2+3(2k^\prime+k)^2}{\Lambda^2_{B_q}}\biggr]
\nonumber\\[5mm]
&\times&\bar u(p_2)\Gamma_2 O^\mu S_{v_1}(k,\bar\Lambda)\Gamma_1 u(v_1)\,\,
{\rm Tr}\bigg[\Gamma_2^\prime (1 + \not\! v_2)
(1 + \not\! v_1)\Gamma_1^\prime \biggr]\nonumber
\end{eqnarray}

\vspace*{1cm}
\underline{Nonfactorizing contributions}

\vspace*{.5cm}
Diagram IIa
\begin{eqnarray}\label{QqIIa-trans}
T^{IIa}_{B_Q\to B_q + M} &=&
\frac{G_F}{\sqrt{2}} \,\, V_{Qq} \,\, V^\dagger_{q_1q_2} \,\,
\frac{N_c!\,\,g_{B_Q}\,\,g_{B_q}\,\,g_M}{(4\pi)^6
\Lambda_q\Lambda_{q_1}\cdots \Lambda_{q_4}}
\int\frac{d^4k}{\pi^2i} \int\frac{d^4k^\prime}{\pi^2i}
\int\frac{d^4k^{\prime\prime}}{\pi^2i}\\[5mm]
&\times&\exp\biggl[\frac{9k^2+3(2k^{\prime\prime}+2k^\prime-k-p_3)^2}
{\Lambda^2_{B_Q}}+\frac{(3k^\prime+2p_2)^2+
3(2k^{\prime\prime}+k^\prime-p_3)^2}{\Lambda^2_{B_q}}\biggr]
\nonumber\\[5mm]
&\times&\bar u(p_2)\Gamma_2 O^\mu S_{v_1}(k,\bar\Lambda)\Gamma_1 u(v_1)\,\,
{\rm Tr}\bigg[\Gamma_2^\prime (1 + \not\! v_2) \Gamma_M
(1 + \not\! v_3) O_\mu (1 + \not\! v_1) \Gamma_1^\prime\biggr]
\nonumber
\end{eqnarray}

\vspace*{0.5cm}
Diagram IIb
\begin{eqnarray}\label{QqIIb-trans}
T^{IIb}_{B_Q\to B_q + M} &=&
\frac{G_F}{\sqrt{2}} \,\, V_{Qq} \,\, V^\dagger_{q_1q_2} \,\,
\frac{N_c!\,\,g_{B_Q}\,\,g_{B_q}\,\,g_M}{(4\pi)^6\Lambda_q
\Lambda_{q_1}\cdots \Lambda_{q_4}}
\int\frac{d^4k}{\pi^2i} \int\frac{d^4k^\prime}{\pi^2i}
\int\frac{d^4k^{\prime\prime}}{\pi^2i}\\[5mm]
&\times&\exp\biggl[\frac{9k^2+3(2k^{\prime\prime}+k+p_3)^2}
{\Lambda^2_{B_Q}}+\frac{(3k^\prime+2p_2)^2+3(2k^{\prime\prime}+
2k-k^\prime+p_3)^2}{\Lambda^2_{B_q}}\biggr]
\nonumber\\[5mm]
&\times&\bar u(p_2)\Gamma_2 O^\mu S_{v_1}(k,\bar\Lambda)\Gamma_1 u(v_1)\,\,
{\rm Tr}\bigg[\Gamma_2^\prime (1 + \not\! v_2) O_\mu \Gamma_M
(1 + \not\! v_3) (1 + \not\! v_1) \Gamma_1^\prime\biggr]
\nonumber
\end{eqnarray}

\vspace*{0.5cm}
Diagram III
\begin{eqnarray}\label{QqIII-trans}
T^{III}_{B_Q\to B_q + M} &=& \frac{G_F}{\sqrt{2}} V_{Qq} V^\dagger_{q_1q_2}
\frac{N_c!\,\,g_{B_Q}\,\,g_{B_q}\,\,g_M}{(4\pi)^6
\Lambda_q\Lambda_{q_1}\cdots \Lambda_{q_4}}
\int\frac{d^4k}{\pi^2i}\int\frac{d^4k^\prime}{\pi^2i}
\int\frac{d^4k^{\prime\prime}}{\pi^2i}\hspace*{0.1cm}\\[5mm]
&\times&\exp\biggl[\frac{9k^2+3(2k^{\prime\prime}-k-p_3)^2}
{\Lambda^2_{B_Q}}+\frac{(3k^\prime+2p_2)^2
+3(2k^{\prime\prime}-k^\prime+p_3)^2}{\Lambda^2_{B_q}}\biggr]\nonumber\\[5mm]
&\times&\bar u(p_2)\Gamma_2 O^\mu S_{v_1}(k,\bar\Lambda)\Gamma_1 u(v_1)\,\,
{\rm Tr}\bigg[\Gamma_2^\prime (1 + \not\! v_2) O_\mu
(1 + \not\! v_1)\Gamma_1^\prime \Gamma_M(1 + \not\! v_3)\biggr]
\nonumber
\end{eqnarray}
\noindent Details of the calculation of the matrix elements
(\ref{BCFD-trans})-(\ref{QqIII-trans}) can be found in Appendix B.

Below we list the Lorentz-spinor parts $\bar u(p_2)...u(p_1) {\rm Tr}[...]$
of the individual diagrams where one has to differentiate between the various
possible light diquark transitions. ($M_1$, $M_2$ and $M_3$ denote the masses
of the initial and final baryons, and the meson, respectively)

\vspace*{1cm}
$\bullet$ {\bf scalar-to-scalar diquark transitions}

\vspace*{.5cm}
Factorizing Diagram  (I)
\begin{eqnarray}
& & (M_1M_2M_3) v_3^\mu[\bar u (\not \! v_2 + 1) O_\mu
(\not \! v_1 + 1) u] {\rm Tr}[\gamma_5 (\not \! v_2 + 1)
(\not \! v_1 + 1) \gamma_5]\nonumber\\
& &\nonumber\\
&=& 8 Q_+ \bar u (M_- - M_+ \gamma_5) u\biggr|_{M_2/M_1\to 0}
\Longrightarrow 8 M_1^3 \bar u (1 - \gamma_5) u
\label{FDPP}
\end{eqnarray}

\newpage
Diagram IIa
\begin{eqnarray}
& & (M_1M_2M_3) [\bar u (\not \! v_2 + 1) O_\mu (\not \! v_1 + 1) u]
{\rm Tr}[\gamma_5(\not \! v_2 + 1) \gamma_5
(\not \! v_3 + 1) O_\mu  (\not \! v_1 + 1) \gamma_5]\nonumber\\
& &\nonumber\\
&=& 16 M_1 \bar u [- P_+ - \gamma_5 Q_+] \biggr|_{M_2/M_1\to 0}
\Longrightarrow  16 M_1^3 \bar u (1 - \gamma_5) u
\label{IIaPP}
\end{eqnarray}

\vspace*{.5cm}
Diagram IIb
\begin{eqnarray}
&-& (M_1M_2M_3) [\bar u (\not \! v_2 + 1) O_\mu (\not \! v_1 + 1) u]
{\rm Tr}[\gamma_5 (\not \! v_2 + 1) O_\mu \gamma_5
(\not \! v_3 + 1) (\not \! v_1 + 1)\gamma_5]\nonumber\\
& &\nonumber\\
&=& 16 M_2 \bar u [D_+ - \gamma_5 Q_+] u \biggr|_{M_2/M_1\to 0}
\Longrightarrow 16 M_1^2M_2 \bar u (1 - \gamma_5) u
\label{IIbPP}
\end{eqnarray}

\vspace*{.5cm}
Diagram III
\begin{eqnarray}
& &(M_1M_2M_3) [\bar u (\not \! v_2 + 1) O_\mu (\not \! v_1 + 1) u]
{\rm Tr}[\gamma_5 (\not \! v_2 + 1) O_\mu
(\not \! v_1 + 1) \gamma_5 \gamma_5 (\not \! v_3 + 1)]\nonumber\\
& &\nonumber\\
&=& 32 \,\,\, (M_1M_2) \,\,\, \sum_{i=1}^3 M_i  \bar u \gamma_5 u
\biggr|_{M_2/M_1\to 0}
\Longrightarrow  32 M_1^2M_2 \bar u \gamma_5 u
\label{IIIPP}
\end{eqnarray}

\vspace*{1cm}
$\bullet$ {\bf vector-to-scalar diquark transitions}

\vspace*{.5cm}
Diagram IIa
\begin{eqnarray}
& &(M_1M_2M_3) [\bar u \gamma^\beta\gamma^5(\not \! v_2 + 1) O_\mu
(\not \! v_1 + 1) u] {\rm Tr}[\gamma_\beta(\not \! v_2 + 1) \gamma_5
(\not \! v_3 + 1) O_\mu  (\not \! v_1 + 1) \gamma_5]\nonumber\\
& &\nonumber\\
&=& 16 M_1 \bar u [ 3P_+ - \gamma_5 Q_+] u \biggr|_{M_2/M_1\to 0}
\Longrightarrow  - 16 M_1^3 \bar u (3 + \gamma_5) u
\label{IIaVP}
\end{eqnarray}

\vspace*{.5cm}
Diagram IIb
\begin{eqnarray}
& &(M_1M_2M_3) [\bar u \gamma^\beta\gamma^5(\not \! v_2 + 1) O_\mu
(\not \! v_1 + 1) u] {\rm Tr}[\gamma_\beta(\not \! v_2 + 1)
O_\mu \gamma_5 (\not \! v_3 + 1) (\not \! v_1 + 1)\gamma_5]\nonumber\\
& &\nonumber\\
&=& - 48 M_2 \bar u [D_+ - \gamma_5 Q_+] u \biggr|_{M_2/M_1\to 0}
\Longrightarrow - 48 M_1^2M_2 \bar u (1 - \gamma_5) u
\label{IIbVP}
\end{eqnarray}

\vspace*{.5cm}
Diagram III
\begin{eqnarray}
& &(M_1M_2M_3) [\bar u \gamma^\beta\gamma^5(\not \! v_2 + 1) O_\mu
(\not \! v_1 + 1) u] {\rm Tr}[\gamma_\beta(\not \! v_2 + 1)
O_\mu (\not \! v_1 + 1) \gamma_5 \gamma_5 (\not \! v_3 + 1)]\nonumber\\
& &\nonumber\\
&=& - 96 (M_1M_2) \sum_{i=1}^3 M_i \hspace*{0.3cm} \bar u \gamma_5 u
\biggr|_{M_2/M_1\to 0} \Longrightarrow -96M_1^2M_2\bar u\gamma_5u
\label{IIIVP}
\end{eqnarray}

\vspace*{1cm}
$\bullet$ {\bf vector-to-vector diquark transitions}

\vspace*{.5cm}
Factorizing Diagram  (I)
\begin{eqnarray}
& & (M_1M_2M_3) v_3^\mu[\bar u \gamma^\alpha\gamma^5
(\not \! v_2 + 1) O_\mu (\not \! v_1 + 1) \gamma^\beta\gamma^5 u]
{\rm Tr}[\gamma_\alpha (\not \! v_2 + 1) (\not \! v_1 + 1) \gamma_\beta]
\nonumber\\
& &\nonumber\\
&=& - 8 Q_+ \bar u (3M_- + M_+ \gamma_5) u \biggr|_{M_2/M_1\to 0}
\Longrightarrow - 8 M_1^3 \bar u (3 + \gamma_5) u
\label{FDVV}
\end{eqnarray}

\vspace*{.5cm}
Diagram IIa
\begin{eqnarray}
& & (M_1M_2M_3) [\bar u \gamma^\alpha\gamma^5(\not \! v_2 + 1)
O_\mu (\not \! v_1 + 1) \gamma^\beta\gamma^5 u]
{\rm Tr}[\gamma_\alpha(\not \! v_2 + 1) \gamma^5 (\not \! v_3 + 1)
O_\mu  (\not \! v_1 + 1) \gamma_\beta]\nonumber\\
& &\nonumber\\
&=&  48 M_1 \bar u [ 3P_+ - \gamma_5 Q_+] u \biggr|_{M_2/M_1\to 0}
\Longrightarrow  - 48 M_1^3 \bar u (3 + \gamma_5) u
\label{IIaVV}
\end{eqnarray}

\vspace*{.5cm}
Diagram IIb
\begin{eqnarray}
& & (M_1M_2M_3) [\bar u \gamma^\alpha\gamma^5(\not \! v_2 + 1)
O_\mu (\not \! v_1 + 1) \gamma^\beta\gamma^5 u]
{\rm Tr}[\gamma_\alpha (\not \! v_2 + 1) O_\mu \gamma_5
(\not \! v_3 + 1) (\not \! v_1 + 1)\gamma_\beta]\nonumber\\
& &\nonumber\\
&=& - 48 M_2 \bar u [3 D_+ + \gamma_5 Q_+]u \biggr|_{M_2/M_1\to 0}
\Longrightarrow -48 M_1^2M_2 \bar u (3 + \gamma_5) u
\label{IIbVV}
\end{eqnarray}

\vspace*{.5cm}
Diagram III
\begin{eqnarray}
& &(M_1M_2M_3) [\bar u \gamma^\alpha\gamma^5(\not \! v_2 + 1)
O_\mu (\not \! v_1 + 1) \gamma^\beta\gamma^5 u]
        [\gamma_\alpha (\not \! v_2 + 1) O_\mu
(\not \! v_1 + 1) \gamma_\beta \gamma^5 (\not \! v_3 + 1)]\nonumber\\
& &\nonumber\\
&=& - 288 \,\,\, (M_1M_2) \,\,\, \sum_{i=1}^3 M_i  \bar u \gamma_5 u
\biggr|_{M_2/M_1\to 0} \Longrightarrow  - 288 M_1^2M_2 \bar u \gamma_5 u
\label{IIIVV}
\end{eqnarray}
\noindent where
$$
Q_+=(M_1+M_2)^2-M_3^2, \,\,\,
P_+=(M_2+M_3)^2-M_1^2, \,\,\,
D_+=(M_1+M_3)^2-M_2^2
$$

The relations Eqs.~(\ref{FDPP})-(\ref{IIIVV}) are in a complete agreement
with the result of spectator model analysis \cite{Kramer}. Note also that
the contributions arising from the diagrams IIb and III can be seen to be
down by the helicity flip factor $(M_2/M_1)$ in agreement with the result
of \cite{Kramer}.

The general invariant matrix element describing exclusive weak nonleptonic
decays of heavy baryons $1/2^+\to 1/2^+ + 0^-$ is given by one

\begin{equation}\label{Minv}
M= M_{\rm I}+M_{\rm IIa}+M_{\rm IIb}+M_{\rm III}\equiv A-\gamma_5 B
\end{equation}
where the amplitudes $M_{\rm I}$, $M_{\rm IIa}$,
$M_{\rm IIb}$, and $M_{\rm III}$
are determined from the  diagrams I, IIa, IIb, and III, respectively.
Our results are given in the form

\noindent
\underline{Factorizing contribution:}
\begin{equation}\label{fac1}
{\rm Diagram \;\; I:} \hspace{.5cm}
M_{\rm I}=c_W\chi_{\pm}f_P\frac{Q_+}{4M_1M_2}
\biggl(M_-\ell^-_{FD}-M_+\ell^+_{FD} \cdot \gamma^5\biggr) f(M_1,M_2,M_3)
\end{equation}

\vspace{0.3cm}
\noindent
\underline{Nonfactorizing contributions}
\begin{eqnarray}
& &{\rm Diagram \;\; IIa:} \hspace{.5cm} M_{\rm IIa}=
c_Wc_-\frac{H_2(M_1,M_2,M_3)}{4M_1M_2}
\biggl(P_+\ell^{P^+}_{II_a}-Q_+\ell^{Q^+}_{II_a} \cdot \gamma^5\biggr) M_1
\label{fac2a}\\
& &\nonumber\\
& &{\rm Diagram \;\; IIb:} \hspace{.5cm}
M_{\rm IIb}=c_Wc_-\frac{H_2(M_1,M_2,M_3)}{4M_1M_2}
\biggl(D_+\ell^{D^+}_{II_b}-Q_+\ell^{Q^+}_{II_b} \cdot \gamma^5 \biggr) M_2
\label{fac2b}\\
& & \nonumber\\
& &{\rm Diagram \;\; III:} \hspace{.5cm}
M_{\rm III}=c_Wc_-\frac{H_3(M_1,M_2,M_3)}{4M_1M_2}
\sum_{i=1}^3M_i(M_1M_2)\ell_{III} \cdot \gamma^5\label{fac3}
\end{eqnarray}

Here,
$c_W=G_F/\sqrt{2}V_{QQ^\prime(q)}V^\dagger_{q_1q_2}$, $f_P$ ($P=\pi$, $K$)
are meson leptonic decay constants;
$c_- = c_1 - c_2$ and $\ell^\pm_{FD}$, $\ell^{P^+}_{II_a}$,
$\ell^{Q^+}_{II_a}$, $\ell^{D^+}_{II_b}$, $\ell^{Q^+}_{II_b}$,
$\ell_{III}$ are flavor coefficients whose values are listed
in Tables IIa and IIb.
The full list of expressions for the form factors $f(M_1,M_2,M_3)$,
$H_2(M_1,M_2,M_3)$ and $H_3(M_1,M_2,M_3)$ appearing in
Eqs.~(\ref{fac1})-(\ref{fac3}) is given below. At the present stage we
only give a complete analysis of the Cabibbo-favored nonleptonic decays
only for $1/2^+ \to 1/2^+ + 0^{-}$ transitions. In addition to these decays
we shall also consider the factorizing processes with vector mesons
$\Lambda_c^+\to p\phi$ and $\Lambda_b^0\to J/\psi \Lambda$ which were
recently measured by the CLEO~\cite{CLEO} and CDF~\cite{CDF} Collaborations.

\newpage
\underline{$b\to c$ transitions}
\begin{eqnarray}\label{BCq2}
f(\omega)&=&\frac{R(\omega,\bar\Lambda)}{R(1,\bar\Lambda)}, \hspace{1cm}
\omega=\frac{M_1^2+M_2^2-M_3^2}{2M_1M_2}\\
H_i(\omega)&=&d_i\,t_i(r)\,
\frac{R_H(\omega,\bar\Lambda^{i},\bar\Lambda^{f})}
{\sqrt{R(1,\bar\Lambda^{i})R(1,\bar\Lambda^{f})}}
\,\frac{8}{9\pi\sqrt{3}}
\frac{\Lambda_{B_Q}^4}{\Lambda^3}
\hspace{1cm} (i=2,3)
\nonumber
\end{eqnarray}
where
\begin{eqnarray}
\hspace*{-.8cm}
R(\omega,\bar\Lambda)&=&\int\limits_0^\infty du u\int\limits_0^1d\alpha
\exp\biggl\{-18u^2 [1+2\alpha(1-\alpha)(\omega-1)]
+36u \bar\Lambda/\Lambda_{B_Q}\biggr\}
\nonumber\\
\hspace*{-.8cm}
R_{H}(\omega,\bar\Lambda^{i},\bar\Lambda^{f})
&=& \int\limits_0^\infty du u\int\limits_0^1d\alpha
\exp\biggl\{-72u^2[1+2\alpha(1-\alpha)(\omega-1)]\biggr\}
\nonumber\\
&+&\exp\biggl\{144u(\bar\Lambda^i\alpha+\bar\Lambda^f(1-\alpha))/
\Lambda_{B_Q}-432u^2(\alpha^2+(1-\alpha)^2)\biggr\}
\nonumber
\end{eqnarray}
Here $d_2=1$ and $d_3=0.5\exp[9M_3^2/2\Lambda_{B_Q}^2]$.
The parameters $\bar\Lambda^i$ and $\bar\Lambda^f$ correspond to initial
and final baryons, respectively. The parameters
$t_i(r)$, where $r=\Lambda/\Lambda_s$, are given in Table IIIa.

It is well-known that there are altogether three IW functions
$\zeta(\omega)$, $\xi_1(\omega)$ and $\xi_2(\omega)$ describing current
induced ground state to ground state transitions. Here $\zeta(\omega)$
describes $\Lambda_Q$-type baryon transitions and $\Omega_Q$-type baryon
transitions \cite{Isgur,Georgi}. In our approach they are expressed via
a single universal function $f(\omega)$
\begin{eqnarray}\label{BarIW}
\zeta(\omega)=\xi_1(\omega)=\xi_2(\omega)(1+\omega)=
f(\omega)\frac{Q_+}{M_1M_2}=f(\omega)\frac{\omega+1}{2},\,\,\,\,\,\,\,\,
f(1)=1
\end{eqnarray}
\noindent This result coincides with the prediction of large-N$_c$ QCD
\cite{Chow} and reproduces the result of the spectator quark model
\cite{DESY}.

\vspace*{1cm}
\underline{Heavy-light transitions}
\begin{eqnarray}\label{hl_trans}
f(M_1,M_2,M_3)&=&\frac{R_{FD}(M_1,M_2,M_3,\bar\Lambda)}
{\sqrt{R(1,\bar\Lambda)}}
\frac{8R^2}{(1+R)^3}\frac{1}{\sqrt{\chi(r)}},\,\,\,\,\,
R=\frac{\Lambda_{B_Q}^2}{\Lambda_{B_q}^2}\\
& &\nonumber\\
R_{FD}(M_1,M_2,M_3,\bar\Lambda)&=&
\hspace*{-.15cm}\int\limits_0^\infty\hspace*{-.1cm}d\alpha
\exp\biggl[ -9\alpha^2(1+R)+18\alpha\frac{\bar\Lambda}{\Lambda_{B_Q}}(1+R)
\biggr]\nonumber\\
&\times&\exp\biggl[ -12\alpha R\omega \frac{M_2}{\Lambda_{B_Q}}+
\frac{4R}{R+1}\frac{M_2^2}{\Lambda_{B_Q}^2}\biggr]
\nonumber\\
& &\nonumber\\
H_i(M_1,M_2,M_3,\bar\Lambda)&=&\frac{t_i(r)}{\sqrt{\chi(r)}}\,
\frac{R_{H_i}(M_1,M_2,M_3,\bar\Lambda)}
{(1+R)\sqrt{R(1,\bar\Lambda)}}
\,\frac{4}{9\pi\sqrt{3}}
\frac{\Lambda_{B_Q}^4}{\Lambda^3}
\hspace{1cm} (i=2,3)
\nonumber
\end{eqnarray}
where
\begin{eqnarray}
R_{H_2}&=&\int\limits_0^\infty d\alpha\exp\biggr[ 36\alpha^2(1+R)(3R+4)
+72\alpha\frac{\bar\Lambda}{\Lambda_{B_Q}}(1+R)
-12\alpha R\omega\frac{M_2}{\Lambda_{B_Q}} +
\frac{R}{1+R}\frac{M_2^2}{\Lambda_{B_Q}^2}\biggr]
\nonumber\\
& &\nonumber\\
R_{H_3}&=&\int\limits_0^\infty d\alpha\exp\biggr[ 36\alpha^2(1+R)(3R+4)
+72\alpha\frac{\bar\Lambda}{\Lambda_{B_Q}}(1+R)
-12\alpha R\frac{M_2^2+M_3^2}{\Lambda_{B_Q}M_1}\biggr]\nonumber\\
& &\nonumber\\
&\times&\exp\biggl[\frac{R}{1+R}
\frac{M_2^2+6M_3^2}{\Lambda_{B_Q}^2}\biggr]
\nonumber
\end{eqnarray}
\noindent The parameters $\chi(r)$, $t_2(r)$ and $t_3(r)$
are given in Table IIIb.
The terms proportional to $(M_1-M_2)/\Lambda_{B_Q}$ in the exponents in
Eqs.~(\ref{BCq2}) and (\ref{hl_trans}) have been dropped for physical
reasons.

\section{Results}

In this section we give our numerical results for the decay rates and
the asymmetry parameters in the nonleptonic decays of $\Lambda_Q$, $\Xi_Q$
and $\Omega_Q$ baryons. Let us specify the model parameters.
Our model contains the following set of parameters: the cutoff parameters
$\Lambda_{B_q}$ and $\Lambda_{B_Q}$ and the binding energy parameters
($\bar\Lambda, \bar\Lambda_s$ and $\bar\Lambda_{ss}$).
Three of the parameters ($\Lambda_{B_q}$, $\Lambda_{B_Q}$ and $\bar\Lambda$)
are used to fit known branching ratios of five nonleptonic decays
$\Lambda^+_c\to\Lambda^0\pi^+$, $\Lambda^+_c\to \Sigma^0\pi^+$,
$\Lambda^+_c\to \Sigma^+\pi^0$, $\Lambda^+_c\to p\bar K^0$
and $\Lambda^+_c\to \Xi^0 K^+$. Moreover, in the fit we impose the
condition $\rho^2 = 1$ on the slope of baryonic Isgur-Wise function.
The fit yields the following values for these model parameters:
$\Lambda_{B_q}$=3.037 GeV, $\Lambda_{B_Q}$=2.408 GeV, $\bar\Lambda$=0.9 GeV.
One has to remark that the values $\Lambda_{B_q}$ and $\Lambda_{B_Q}$ are
the phenomenological parameters which in principle are related to the size
of a baryon. However their magnitude is not strongly constrained by the
experimental values of baryon observables and allows for the variation in a
rather wide range. Note that the obtained value $\Lambda_{B_Q}$=2.408 GeV is
close to $\Lambda_{B_Q}$ = 2.5 GeV coming from analysis of semileptonic heavy
baryon decays in relativistic three-quark model which uses the constituent
quark masses \cite{Mainz}. As to the cutoff parameter in the light-baryon
vertex, in Ref.~\cite{PSI} it was demonstrated that the experimental data
both for the dimensionless (nucleon magnetic moments) as well as
dimensionful (nucleon charge radii) observables can be described
successfully, using the value of the parameter $\Lambda_{B_q}$ from the
interval $\sim$ (1--3) GeV provided the constituent quark mass is properly
fitted. In particular, for the value $\Lambda_{B_q}$ = 3.037 GeV,
with the constituent quark mass $m_q$ = 315 MeV, we obtain for the nucleon
magnetic moments and charge radii:
$\mu_p$ = 2.62 (experiment 2.79), $\mu_n$ = - 1.61 (experiment -1.91),
$r^E_p$ = 0.82 fm (experiment 0.86$\pm$0.01 fm),
$<r^2>^E_n$ = -0.188 fm$^2$ (experiment -0.119$\pm$0.004 fm$^2$),
$r^M_p$ = 0.74 fm (experiment 0.86$\pm$0.06 fm),
$r^M_n$ = 0.76 fm (experiment 0.88$\pm$0.07 fm).
The parameters $\bar\Lambda_s$
and $\bar\Lambda_{ss}$ cannot be determined at present due to the lack
of experimental information on the decays of heavy-light baryons containing
one or two strange quarks. For the time being we fix them at the values
$\bar\Lambda_s$=1 GeV and $\bar\Lambda_{ss}$=1.1 GeV.
The masses of hadrons are taken from \cite{PDG,DESY}.
In what follows we will use the following values for the
Cabibbo-Kobayashi-Maskawa matrix elements $V_{qq^\prime}$ \cite{PDG}:
\begin{eqnarray}
|V_{cb}| = 0.04, \,\,\,\,\, |V_{ud}| \approx |V_{cs}| = 0.975,
\,\,\,\,\, |V_{us}| \approx |V_{cd}| = 0.22, \,\,\,\,\,
|V_{ub}| = 0.0035
\end{eqnarray}
\noindent The Wilson coefficients are taken to be
$a_1$ = 1.03, $a_2$ = 0.10, $a_1^\star$ = 1.3, $a_2^\star$ = -0.65.

In order to check on the consistency of our approach, we shall prove that the
Isgur-Wise functions $\xi_1$ and $\xi_2$ satisfy the model-independent
Bjorken-Xu inequalities \cite{Xu}.
As was mentioned in Sec. 3 the baryonic IW functions
$\zeta(\omega)$, $\xi_1(\omega)$ and $\xi_2(\omega)$, corresponding
to $\Lambda_Q$-type and $\Omega_Q$-type heavy-heavy weak baryon transitions,
are expressed via a single universal function $f(\omega)$
(see, Eqs.~(\ref{BCq2}) and \ref{BarIW})).

The IW-functions $\xi_1$ and $\xi_2$ must satisfy to the two
model-independent Bjorken-Xu inequalities in \cite{Xu}. The first inequality
reads
\begin{eqnarray}\label{ineq1}
1\geq \frac{2+\omega^2}{3}\xi_1^2(\omega)+
\frac{(\omega^2-1)^2}{3}\xi_2^2(\omega)
+\frac{2}{3}(\omega-\omega^3)\xi_1(\omega)\xi_2(\omega)
\end{eqnarray}
\noindent The inequality (\ref{ineq1}) implies a second inequality,
namely a model-independent restriction on the slope (radius) of the
form factor $\xi_1(\omega)$
\begin{eqnarray}\label{ineq2}
\rho^2_{\xi_1}\equiv - \left.\frac{d\xi_1(\omega)}{d\omega}\right|_{\omega=1}
\geq \frac{1}{3}-\frac{2}{3}\xi_2(1)
\end{eqnarray}

From the inequality (\ref{ineq1}) we find an upper limit for the
universal function $f(\omega)$
\begin{eqnarray}\label{ineq3}
\xi_1(\omega) \leq 1 \,\,\, \text{or}\,\,\,
f(\omega) \leq \sqrt{\frac{2}{1+\omega}}
\end{eqnarray}
\noindent which we impose as a condition.

From the inequality (\ref{ineq2}) for the slope of the function
$\xi_1(\omega)$ we see that $\rho^2_{\xi_1} \geq 0$.
For the choice of model parameters corresponding to {\it the best fit}
the universal function $f(\omega)$ and the slope of the $\xi_1$ satisfy to
the Bjorken-Xu inequalities (\ref{ineq1}) and (\ref{ineq2}). In this case
the charge radii of the $\zeta$ and $\xi$ functions are equal to 0.84.
Our form factor function $f(\omega)$ is well approximated by the formula
\begin{eqnarray}
f(\omega) \approx \biggl[\frac{2}{1+\omega}\biggr]^{1 + 0.68 / \omega}
\end{eqnarray}

In Table IV we present the branching ratios of the decays
$\Lambda^+_c\to\Lambda^0\pi^+$, $\Lambda^+_c\to \Sigma^0\pi^+$,
$\Lambda^+_c\to \Sigma^+\pi^0$, $\Lambda^+_c\to p\bar K^0$ and
$\Lambda^+_c\to \Xi^0 K^+$ which are described nicely using a
three-parameter fit. Our predictions for the other heavy-to-light
decay modes are listed in Table IV. In Table V we give the calculated values
for the asymmetry parameters in the nonleptonic decays of $1/2^+$ charm
and bottom baryons into octet of light baryons and pseudoscalar mesons
(pions and kaons). The relevant formulae for the decay rates and the
asymmetry parameters in terms of the invariant
amplitudes $A$ and $B$ are listed in ref. \cite{Kramer}.
For comparison in Tables IV and V we quote the results predicted by other
phenomenological approaches. It is seen that rates of decays which proceed
only via the nonfactorizing diagrams are not suppressed.
In Table VI we list our predictions for the parity-violating ($A$) and
parity-conserving ($B$) amplitudes in the decays
$\Lambda_c^+\to \Lambda\pi^+$ and $\Lambda_c^+\to \Sigma^+\pi^0$ in units of
$G_FV_{cs}V_{ud}\times $10$^{-2}$ GeV$^2$.

In Table VII we give the predictions for the rates and the asymmetry
parameters in the nonleptonic decays of bottom baryons into charm
baryons with the use of the same model parameters. A clear pattern emerges.
The dominant rates are into channels with factorizing contributions.
Rates which proceed only via nonfactorizing diagrams are small but not
negligibly small.
The total contribution of the nonfactorizing diagrams can be seen to be
destructive. The sum of nonfactorizing contributions amount
up to 30~\% of the factorizing contribution in amplitude.
Using $\tau(\Lambda_b)=(1.14 \pm 0.08) \times 10^{-12}$ s \cite{PDG}
we predict a branching ratio of the mode $\Lambda_b\to\Lambda_c \pi$ to be
$(0.44\pm 0.003)\%$. If one neglects the nonfactorizing contributions
for this mode as was done in~\cite{Cheng3} one would obtain an enhanced
rate of $\Gamma=0.665\times 10^{10} {\rm s}^{-1}$. The prediction for
the asymmetry parameter remains at $\alpha\simeq -1$ and is thus not
affected by such an omission.

In Tables VIII and IX we analyze the nonfactorizing contributions to
the decay amplitudes for the transitions
$\Lambda_c^+\to \Lambda\pi^+$ and $\Lambda_b^0\to \Lambda^+_c\pi^-$.
It is seen that the total contribution of the nonfactorizing diagrams are
destructive. They can amount up to $\sim 60~\%$ of the factorizing
contribution in amplitude of heavy-to-light transition and up to $\sim 30~\%$
of the factorizing contribution in amplitude of $b\to c$ transition. Also
we calculate the values for overlap integrals $f$, $H_2$ and $H_3$ for these
modes. They turn out to be equal to $f=0.51$, $H_2$=43 MeV and $H_3$=14 MeV
for $\Lambda_c^+\to \Lambda\pi^+$ and $f=0.61$, $H_2$=24 MeV and $H_3$=12 MeV
for $\Lambda_b^0\to \Lambda^+_c\pi^-$. For comparison we quote the results
for overlap integrals evaluated for the decay  $\Lambda_c^+\to \Lambda\pi^+$
in ref.~\cite{Kramer} : $f=0.34$, $H_2$=40 MeV and $H_3$=-4 MeV.

In Tables X and XI we present the predictions for the $\Lambda^+_c\to p\phi$
and $\Lambda_b^0\to J/\psi\Lambda$ decays for various values of the $a_2$ and
$a^\star_2$ parameters. As mentioned before these processes are described by
the factorizing diagram alone. The corresponding weak hadronic matrix
elements in the spectator approximation have a trivial spin structure given
by the matrix $O_\mu$. For this reason the asymmetry parameter for these
transitions does not depend on the model parameters and can be expressed
through the hadron masses
\begin{eqnarray}
\alpha\left(\frac{1}{2}^+\rightarrow\frac{1}{2}^++1^-\right)=
-\frac{M_1^2-M_2^2-2M_3^2}
{\sqrt{Q_+Q_-}+\frac{3}{2}M_3^2(Q_+ + Q_-)}
\end{eqnarray}
\noindent In particular, the asymmetry parameter in the decay
$\Lambda^+_c\to p\phi$ is equal to $-0.26$ and  $\alpha=0.21$ for
$\Lambda_b^0\to J/\psi\Lambda$ transition.
It is seen that for the accepted value of the Wilson coefficient $a_2=0.10$
our approach gives the prediction for the branching
$Br(\Lambda_b^0\to J/\psi \Lambda)$ = 0.027 which is
consistent with the recent CDF data
$Br(\Lambda_b^0\to J/\psi\Lambda)$ = 0.037$\pm$0.017$\pm$0.004 \cite{CDF}.
For the rare decay $\Lambda_c^+\to p \phi$ our approach for
the accepted value of the corresponding Wilson coefficient
$a^\star_2$ = -0.65 yields the branching ratio
$Br(p\phi)/Br(pK^-\pi^+)=0.105$ which overestimates the known experimental
data from CLEO \cite{CLEO} and NA32 \cite{NA32} measurements.

\section{Conclusion}
We have studied the exclusive nonleptonic decays of heavy-light baryons
into charm and light baryons. The decay rates and the asymmetry parameters
have been calculated. It would be interesting to test our predictions
in $b\to c$ transitions in future experiments.

We have shown that rates of decays which proceed only via the nonfactorizing
diagrams are suppressed but not completely suppressed for both cases of
heavy-to-light and heavy-to-heavy transitions. We have analyzed in detail the
nonfactorizing contributions to the decay amplitudes for the transitions
$\Lambda_c^+\to \Lambda\pi^+$ and $\Lambda_b^0\to \Lambda^+_c\pi^-$. It was
shown that the total contribution of the nonfactorizing diagrams are
destructive. They amount up to $\sim 60~\%$ of the factorizing
contribution in amplitude of heavy-to-light transition and up to $\sim 30~\%$
of the factorizing contribution in amplitude of $b\to c$ transition.
Finally, we give the predictions for the $\Lambda^+_c\to p\phi$ and
$\Lambda_b^0\to J/\psi\Lambda$ decays for various values of the $a_2$ and
$a^\star_2$ parameters.

The generalization to the channels $\frac{1}{2}^+\to\frac{1}{2}^++1^-$,
$\frac{1}{2}^+\to\frac{3}{2}^++0^-$ and $\frac{1}{2}^+\to\frac{1}{2}^++1^-$
involving the ground state partners of the mesons and baryons in the final
state is straightforward and will be treated in a subsequent paper.
In this paper we have only discussed the Cabibbo favoured decays induced
by the transitions $b\to c\bar u d$ with a light pseudoscalar meson in
the final state. There are also a number of Cabibbo favoured decays with
heavy mesons in the final state which include the decays induced by the
quark transitions $ b\to c \bar c s$. The treatment of heavy mesons in the
final state requires some refinements in our simple Lagrangian spectator
model. Again, exclusive nonleptonic heavy baryon decays involving heavy
mesons in the final state are the subject of a future publication.

\acknowledgments

M.A.I, V.E.L and A.G.R thank Mainz University for the hospitality
where a part of this work was completed.
This work was supported in part by the Heisenberg-Landau Program,
by the Russian Fund of Basic Research (RFBR) under contract
96-02-17435-a, the State Committee of the Russian Federation for
Education (project N 95-0-6.3-67) and by the BMBF (Germany) under
contract 06MZ566. J.G.K. acknowledges partial support by the BMBF
(Germany) under contract 06MZ566.

\newpage

\appendix

\section{HADRON-QUARK INTERACTION LAGRANGIANS}

\setcounter{equation}{0}
\def\theequation{A\arabic{equation}}

\vspace*{0.5cm}
Below we present a complete list of hadronic interaction Lagrangians used
in the calculations.
We start from the consideration of various possible couplings of three quarks
in the light baryons. It is well known that there are five possible
nonderivative forms of such coupling for octet baryons \cite{Diquark}
\begin{eqnarray}\label{forms}
\begin{array}{ll}
\hspace*{.3cm}\text{\it Pseudoscalar variant}
& \bar B^{km_1} q^{a_1}_{m_1} q^{a_2}_{m_2} C\gamma^5 q^{a_3}_{m_3}
\varepsilon^{a_1a_2a_3} \varepsilon^{km_2m_3}\\[5mm]
\hspace*{.3cm} \text{\it Scalar variant}
& \bar B^{km_1} \gamma^5q^{a_1}_{m_1} q^{a_2}_{m_2} C q^{a_3}_{m_3}
\varepsilon^{a_1a_2a_3} \varepsilon^{km_2m_3}\\[5mm]
\hspace*{.3cm} \text{\it Axial variant}
& \bar B^{km_1} \gamma^\mu q^{a_1}_{m_1} q^{a_2}_{m_2} C
\gamma_\mu\gamma_5 q^{a_3}_{m_3}
\varepsilon^{a_1a_2a_3} \varepsilon^{km_2m_3}\\[5mm]
\hspace*{.3cm} \text{\it Vector variant}
& \bar B^{km} \lambda^{mm_1}_i \gamma^\mu\gamma^5 q^{a_1}_{m_1}
q^{a_2}_{m_2} \lambda^{nm_3}_i C \gamma_\mu q^{a_3}_{m_3}
\varepsilon^{a_1a_2a_3} \varepsilon^{km_2n}\\[5mm]
\hspace*{.3cm} \text{\it Tensor variant}
& \bar B^{km} \lambda^{mm_1}_i \sigma^{\mu\nu}\gamma^5 q^{a_1}_{m_1}
q^{a_2}_{m_2} \lambda^{nm_3}_i C \sigma_{\mu\nu} q^{a_3}_{m_3}
\varepsilon^{a_1a_2a_3} \varepsilon^{km_2n}\\[5mm]
\end{array}
\end{eqnarray}
\noindent where $\bar B^{km}$ is the baryonic octet matrix
\begin{eqnarray}
\bar B^{km}=
\left(
\begin{array}{ccc}
{\bar{\Sigma}^0}/{\sqrt{2}}+{\bar{\Lambda}^0}/{\sqrt{6}}
 & \bar\Sigma^- & -\bar\Xi^- \\
\bar\Sigma^+ & -{\bar\Sigma^0}/{\sqrt{2}}+{\bar\Lambda^0}/{\sqrt{6}}
& \bar\Xi^0 \\
\bar p & \bar n & -{2\bar\Lambda^0}/{\sqrt{6}}\\
\end{array}
\right)
\end{eqnarray}
\noindent It is well known~\cite{Lubovit,Diquark} that these five forms
can be combined in the two linearly independent SU(3) invariant combinations
called {\it vector variant} and {\it tensor variant}
(see, Eqs.~(\ref{currents_l})).

In order to reproduce the results of the spectator model in the Lagrangian
formulation, one has to modify the baryonic currents writing them in terms
of the "projected" quark fields, replacing $q \to V_+ q$,
where $V_+=1/2\,\,(\not\! v+1)$ is the projector introduced is Sec. 2.
With the use of the
"on-shell" conditions $\bar B V_+= \bar B$ and $v^2=1$ it is easy to verify
that there exist simple relations between various interaction Lagrangians
obtained from Eq.~(\ref{forms}) via the  substitution $q \to V_+ q$:
\begin{eqnarray}\label{relations}
& &\bar B^{km_1} V_+ q^{a_1}_{m_1} q^{a_2}_{m_2} C\gamma^5 V_+ q^{a_3}_{m_3}
\varepsilon^{a_1a_2a_3} \varepsilon^{km_2m_3}
= - \bar B^{km_1} \gamma^\mu V_+ q^{a_1}_{m_1} q^{a_2}_{m_2} C
\gamma_\mu\gamma_5 V_+ q^{a_3}_{m_3}
\varepsilon^{a_1a_2a_3} \varepsilon^{km_2m_3}\nonumber\\
& &\bar B^{km_1} \gamma_5 V_+ q^{a_1}_{m_1}
q^{a_2}_{m_2} C V_+ q^{a_3}_{m_3} \varepsilon^{a_1a_2a_3}
\varepsilon^{km_2m_3} = 0\\
& &\bar B^{km} \lambda^{mm_1}_i \gamma^\mu\gamma^5 V_+ q^{a_1}_{m_1}
q^{a_2}_{m_2} \lambda^{nm_3}_i C \gamma_\mu V_+ q^{a_3}_{m_3}
\varepsilon^{a_1a_2a_3} \varepsilon^{km_2n}=\nonumber\\
&=& \frac{1}{2}\bar B^{km} \lambda^{mm_1}_i
\sigma^{\mu\nu}\gamma^5 V_+ q^{a_1}_{m_1}
q^{a_2}_{m_2} \lambda^{nm_3}_i C \sigma_{\mu\nu} V_+ q^{a_3}_{m_3}
\varepsilon^{a_1a_2a_3} \varepsilon^{km_2n}\nonumber
\end{eqnarray}

Since the {\it vector} and {\it tensor} Lagrangians
(\ref{currents_l}) are completely equivalent to each other
on the baryon mass shell one can start with either of them.
Note that the {\it vector} and {\it pseudoscalar}
forms of interaction Lagrangians transform into each other under Fierz
transformations (on baryon mass shell)
\begin{eqnarray}
& &[\bar B^{\alpha_1} V_+^{\alpha_1\alpha_2} q^{\alpha_2}] \otimes
[q^{\alpha_3} (C\gamma_5 V_+)^{\alpha_3\alpha_4} q^{\alpha_4}] =
\frac{1}{2} \biggl\{[\bar B^{\alpha_1} V_+^{\alpha_1\alpha_4}
q^{\alpha_4}] \otimes
[q^{\alpha_3} (C\gamma_5 V_+)^{\alpha_3\alpha_2} q^{\alpha_2}\nonumber\\
&+& [\bar B^{\alpha_1} (\gamma^\mu\gamma^5 V_+)^{\alpha_1\alpha_4}
q^{\alpha_4}] \otimes [q^{\alpha_3} (C\gamma_\mu V_+)^{\alpha_3\alpha_2}
q^{\alpha_2}]\biggr\}
\end{eqnarray}
\noindent Here $(\alpha_i)$ denote the spinor indices.

For SU(3) octet of light baryons the interaction Lagrangians are listed
in Table XII. The interaction Lagrangians for heavy-light baryons are given
in Table XIII. The meson-quark-antiquark interaction Lagrangians are listed
in Table XIV.

The baryon-quark couplings $g_{B}$ are determined from
the normalization condition for vector current.
For heavy-light baryons they are given by
\begin{eqnarray}
g^{-2}_{B_Q} = \frac{N_c!}{(4\pi)^4}
\frac{\Lambda^6_{B_Q}}{18\Lambda_{q_1}\Lambda_{q_2}} \cdot R_Q
\end{eqnarray}
\noindent where $\Lambda_{q_1}$ and $\Lambda_{q_2}$ are light quark
cutoff parameters and $R_Q$ is the structure integral which depends
on the ratio $\bar\Lambda/\Lambda_{B_Q}$
\begin{eqnarray}
R_Q=\int\limits_0^\infty duu
\exp\biggl[-18u^2+36u\frac{\bar\Lambda}{\Lambda_{B_Q}}\biggr]
\nonumber
\end{eqnarray}
\noindent In the case of light baryons the couplings are given by
\begin{eqnarray}
g^{-2}_{B_q} &=& \frac{N_c!}{(4\pi)^4}
\frac{\Lambda^8_{B_q}}{27\Lambda_q^4}\cdot\kappa\\
\text{where}\hspace*{1cm}\kappa&=&\left\{
\begin{array}{cl}
1 & \text{for nucleons}\\
r (2/3 + r/3)
& \text{for $\Lambda^0$ and the triplet of $\Sigma$ hyperons}\\
r^2 (2r/3 + 1/3)
& \text{for the doublet of $\Xi$ hyperons}\\
\end{array}
\nonumber
\right.
\end{eqnarray}

\newpage

\section{THE CALCULATION TECHNIQUE}

\setcounter{equation}{0}
\def\theequation{B\arabic{equation}}

\vspace*{0.5cm}
To elucidate the calculation of the matrix elements
(\ref{BCFD-trans})-(\ref{QqIII-trans})
we consider the four relevant integrals in
Euclidean space
corresponding to the factorizing contributions from $b\to c$
and heavy-light transitions and the typical nonfactorizing ones coming
from the diagram II$_a$. The calculations of the nonfactorizing
contributions from diagrams II$_b$ and III can be carried out analogously.

\vspace*{.5cm}
Factorizing Contribution  ($b\to c$ transition)
\begin{eqnarray}
I_F^{b\to c}(\omega_E)=
\int\frac{d^4 k_E}{\pi^2}\int\frac{d^4 k^\prime_E}{\pi^2}
\exp\biggl[-\frac{18k^2_E+6(2k^{\prime}_E+k)^2}{\Lambda_{B_Q}^2}
\biggr]\frac{1}{k_Ev_{1E}-\bar\Lambda}\,\,
\frac{1}{k_Ev_{2E}-\bar\Lambda}
\end{eqnarray}

\vspace*{.5cm}
Factorizing Contribution ($Q\to q$ transition)
\begin{eqnarray}
I_F^{Q\to q}(\omega_E, M_2)&=&
\int\frac{d^4 k_E}{\pi^2}\int\frac{d^4 k^\prime_E}{\pi^2}
\exp\biggl[-\frac{9k^2_E+3(2k^{\prime}_E+k_E)^2}{\Lambda_{B_Q}^2}
\biggr]\nonumber\\[4mm]
&\times&\exp\biggl[-\frac{(3k_E+2p_{2E})^2+3(2k^{\prime}_E+k_E)^2}
{\Lambda_{B_Q}^2}\biggr]\,\,\frac{1}{k_Ev_{1E}-\bar\Lambda}
\end{eqnarray}

\vspace*{.5cm}
Nonfactorizing Contribution ($b\to c$ transition)
\begin{eqnarray}
I_{NF}^{b\to c}(\omega_E,M_2)&=&
\int\frac{d^4 k_E}{\pi^2}\int\frac{d^4 k^\prime_E}{\pi^2}
\int\frac{d^4 k^{\prime\prime}_E}{\pi^2}
\exp\biggl[-\frac{9k^2_E + 3(2k^{\prime\prime}_E+2k^\prime_E-k_E-p_{3E})^2}
{\Lambda_{B_Q}^2}\biggr]\nonumber\\[4mm]
&\times&\exp\biggl[\frac{9k^{\prime 2}_E+
3(2k^{\prime\prime}_E+k^\prime_E-p_{3E})^2}{\Lambda_{B_Q}^2}
\biggr]\frac{1}{k_Ev_{1E}-\bar\Lambda}\,\,
\frac{1}{k_E^\prime v_{2E}-\bar\Lambda}
\end{eqnarray}

\vspace*{.5cm}
Nonfactorizing Contribution ($Q\to q$ transition)
\begin{eqnarray}
I_{NF}^{Q\to q}(\omega_E,M_2)&=&
\int\frac{d^4 k_E}{\pi^2}\int\frac{d^4 k^\prime_E}{\pi^2}
\int\frac{d^4 k^{\prime\prime}_E}{\pi^2}
\exp\biggl[-\frac{9k^2_E+3(2k^{\prime\prime}_E+2k^\prime_E-k_E-p_{3E})^2}
{\Lambda_{B_Q}^2}\biggr]\nonumber\\[4mm]
&\times&\exp\biggl[-\frac{(3k_E^\prime+2p_{2E})^2+
3(2k^{\prime\prime}_E+k_E-p_{3E})^2}{\Lambda_{B_Q}^2}\biggr]
\,\,\frac{1}{k_Ev_{1E}-\bar\Lambda}
\end{eqnarray}
\noindent The final light baryon state carries the Euclidean momenta
$p_{2E}$ with the mass-shell condition: $p^2_{2E}=-M^2_2$.
The dimensionless variable $\omega_E$ is defined as
$\omega_E=v_{1E}\cdot p_{2E}/M_2 = - \omega$.

Scaling all momentum variables in the above integrals
by $\Lambda_{B_Q}$ and using the Feynman parametrization
\begin{eqnarray}
\frac{1}{A}=\int\limits_0^\infty d\alpha \exp(-\alpha A)
\end{eqnarray}
we have
\begin{eqnarray}\label{Feynman}
I_F^{b\to c}(\omega_E)&=&4\Lambda^6_{B_Q}\int\limits_0^\infty
d\alpha\int\limits_0^\infty  d\beta \int\frac{d^4 k_E}{\pi^2}
\int\frac{d^4 k^\prime_E}{\pi^2}
\exp\biggl[-18k^2_E-24k^{\prime 2}_E\biggr]\nonumber\\[4mm]
&\times& \exp\biggl[-\frac{(\alpha + \beta)^2}{18}
+ \frac{\alpha\beta}{9} (\omega_E + 1) + 2(\alpha + \beta)
\frac{\bar\Lambda}{\Lambda_{B_Q}}\biggr]\nonumber\\[4mm]
I_F^{Q\to q}(\omega_E,M_2)&=&2\Lambda^7_{B_Q}
\int\limits_0^\infty d\alpha\int\frac{d^4 k_E}{\pi^2}
\int\frac{d^4 k^\prime_E}{\pi^2}
\exp\biggl[-9(1+R)k^2_E-12(1+R)k^{\prime 2}_E\biggr] \nonumber\\[4mm]
&\times&\exp\biggl[-\frac{\alpha^2-12R\alpha\omega_E-36RM^2_2}{9(1+R)}
+ 2\alpha\frac{\bar\Lambda}{\Lambda_{B_Q}}\biggr]
\nonumber\\[4mm]
I_{NF}^{b\to c}(\omega_E)&=&4\Lambda^{10}_{B_Q}\int\limits_0^\infty
d\alpha\int\limits_0^\infty d\beta \int\frac{d^4 k_E}{\pi^2}
\int\frac{d^4 k^\prime_E}{\pi^2}\int\frac{d^4 k^{\prime\prime}_E}{\pi^2}
\exp\biggl[-12k^2_E-21k^{\prime 2}_E\biggr]\nonumber\\[4mm]
&\times&\exp\biggl[- \frac{72}{7}k^{\prime\prime 2}_E -
\frac{(\alpha+\beta)^2}{72} + \frac{\alpha\beta}{72} (\omega_E + 1)
+ 2(\alpha+\beta) \frac{\bar\Lambda}{\Lambda_{B_Q}}
- \frac{\alpha^2+\beta^2}{12}\biggr]\nonumber\\[4mm]
I_{NF}^{Q\to q}(\omega_E,M_2)&=&2\Lambda^{11}_{B_Q}
\int\limits_0^\infty d\alpha \int\frac{d^4 k_E}{\pi^2}
\int\frac{d^4 k^\prime_E}{\pi^2}\int\frac{d^4 k^{\prime\prime}_E}{\pi^2}
\exp\biggl[-12k^2_E-3R(3+4R)k^{\prime 2}_E\biggr]\nonumber\\[4mm]
&\times&\exp\biggl[- 36\frac{R(1+R)}{3+4R}k^{\prime\prime 2}_E
-\frac{\alpha^2-12R\alpha\omega_E-9M^2_2}{9(1+R)} +
2\alpha\frac{\bar\Lambda}{\Lambda_{B_Q}}\biggr]
\nonumber
\end{eqnarray}

After integration over $k_E$, $k^\prime_E$ and $k^{\prime\prime}_E$
we arrive at
\begin{eqnarray}\label{arrive}
I_F^{b\to c}(-w) &=& \frac{\Lambda^6_{B_Q}}{12^2}
\int\limits_0^\infty duu \int\limits_0^1 dx
\exp\biggl[ - 18u^2 - 36u^2 x (1 - x) (\omega - 1) +
36 u \frac{\bar\Lambda}{\Lambda_{B_Q}}\biggr]\nonumber\\[4mm]
I_F^{Q\to q}(-M_2, -w) &=&
\frac{2\Lambda^7_{B_Q}}{36^2(1+R)^3}
\int\limits_0^\infty du \exp\biggl[-9(1+R)u^2 +
18(1+R)u\frac{\bar\Lambda}{\Lambda_{B_Q}}\biggr]\nonumber\\[4mm]
&\times&\exp\biggl[-12Ru\omega\frac{M_2}{\Lambda_{B_Q}}+
\frac{4R}{R+1}\frac{M_2^2}{\Lambda^2_{B_Q}}\biggr]\nonumber\\[4mm]
I_{NF}^{b\to c}(-w) &=& \frac{\Lambda^{10}_{B_Q}}{36^2}
\int\limits_0^\infty duu \int\limits_0^1 dx
\exp\biggl[ - 72u^2 - 144u^2 x (1 - x) (\omega - 1)\biggr]
\nonumber\\[4mm]
&\times&\exp\biggl[ 144 u \frac{\bar\Lambda}{\Lambda_{B_Q}}
- 432 u^2 (x^2 + (1-x)^2) \biggr]\nonumber\\[4mm]
I_{NF}^{Q\to q}(-M^2_2, -w) &=& \frac{2\Lambda^{11}_{B_Q}}{216^2R^2(1+R)}
\int\limits_0^\infty du \exp\biggl[ - 36(1+R)(3R+4)u^2\biggr]
\nonumber\\[4mm]
&\times&\exp\biggl[72(1+R)u\frac{\bar\Lambda}{\Lambda_{B_Q}}-
12Ru\omega\frac{M_2}{\Lambda_{B_Q}}+
\frac{R}{R+1}\frac{M_2^2}{\Lambda^2_{B_Q}}\biggr]
\nonumber
\end{eqnarray}

\newpage

\centerline{\bf List of Tables}
\noindent
{\bf TABLE I} Quantum numbers of heavy-light baryons.

\vspace*{.2cm}
\noindent
{\bf TABLE IIa}
Flavor coefficients for heavy-heavy decays
($C\equiv \cos\delta_P$, $S\equiv \sin\delta_P$,
$\delta_P=\theta_P-\theta_I$,
where $\theta_P=-11^{\rm o}$ is the $\eta-\eta^\prime$ mixing angle).

\vspace*{.2cm}
\noindent
{\bf TABLE IIb}
Flavor coefficients for heavy-light decays
($C\equiv \cos\delta_P$, $S\equiv \sin\delta_P$,
${\rm tg}_\pm = 1 \pm {\rm tg}\delta_P \cdot r\sqrt{2}$,
${\rm ctg}_\pm = 1 \pm {\rm ctg}\delta_P \cdot r\sqrt{2}$,
$\delta_P=\theta_P-\theta_I$,
where $\theta_P=-11^{\rm o}$ is
the $\eta-\eta^\prime$ mixing angle, $\theta_I=35^{\rm o}$).

\vspace*{.2cm}
\noindent
{\bf TABLE IIIa}
Factors $t_i(r)$ for heavy-heavy decays
($C\equiv \cos\delta_P$, $S\equiv \sin\delta_P$,
$\delta_P=\theta_P-\theta_I$, where $\theta_P=-11^{\rm o}$ is
the $\eta-\eta^\prime$ mixing angle, $\theta_I=35^{\rm o}$).

\vspace*{.2cm}
\noindent
{\bf TABLE IIIb}
Factors $\chi(r)$ and $t_i(r)$ for heavy-light decays.

\vspace*{.2cm}
\noindent
{\bf TABLE IV} Branching ratios (in $\%$) in nonleptonic
decays $1/2^+\to 1/2^+ + 0^-$ of heavy baryons (heavy-light transitions).
Numerical values of CKM elements and Wilson coefficients:
$|V_{cs}|=0.975$, $|V_{ud}|=0.975$,  $|V_{ub}|=0.0035$,
$a_1^\star=1.3$, $a_2^\star=-0.65$.

\vspace*{.2cm}
\noindent
{\bf TABLE V}
Asymmetry parameters $\alpha$ in the nonleptonic decays
$1/2^+\to 1/2^+ + 0^-$ of heavy baryons (heavy-light transitions).
Numerical values of CKM elements and Wilson coefficients:
$|V_{cs}|=0.975$, $|V_{ud}|=0.975$,  $|V_{ub}|=0.0035$,
$a_1^\star=1.3$, $a_2^\star=-0.65$.

\vspace*{.2cm}
\noindent
{\bf TABLE VI}
Invariant amplitudes in the decay $\Lambda_c^+\to \Lambda\pi^+$
and $\Lambda_c^+\to \Sigma^+\pi^0$
(in units of $G_FV_{cs}V_{ud}\times $10$^{-2}$ GeV$^2$)

\vspace*{.2cm}
\noindent
{\bf TABLE VII}
Decay rates and asymmetry parameters in heavy-heavy transitions.
Numerical values of CKM elements and Wilson coefficients:
$|V_{cb}|=0.04$, $|V_{ud}|=0.975$, $a_1=1.03$, $a_2=0.10$.

\vspace*{.2cm}
\noindent
{\bf TABLE VIII}
Decay $\Lambda_c^+\to \Lambda\pi^+$:
contribution of nonfactorizing diagrams
(in $\%$ relative to the factorizing contribution)

\vspace*{.2cm}
\noindent
{\bf TABLE IX}
Decay $\Lambda_b^0\to \Lambda_c^+\pi^-$:
contribution of nonfactorizing diagrams
(in $\%$ relative to the factorizing contribution)

\vspace*{.2cm}
\noindent
{\bf TABLE X}
Predictions for $\Lambda^+_c\to p\phi$ decay rate
for different values of the Wilson coefficient $a_2^\star$.

\vspace*{.2cm}
\noindent
{\bf TABLE XI}
Predictions for $\Lambda_b^0\to J/\psi\Lambda$ decay rate
for different values of the Wilson coefficient $a_2$.

\vspace*{.2cm}
\noindent
{\bf TABLE XII}
Light baryon Lagrangians

\vspace*{.2cm}
\noindent
{\bf TABLE XIII}
Heavy-light baryon Lagrangians

\vspace*{.2cm}
\noindent
{\bf TABLE XIV}
Meson Lagrangians.

\vspace*{2cm}
\centerline{\bf List of Figures}

\noindent
{\bf FIG. 1}
Diagrams contributing to the matrix element of heavy baryon
nonleptonic decay: factorizing diagram (I),
nonfactorizing diagrams (IIa), (IIb) and (III).

\newpage
\begin{center}
{\bf TABLE I}
\end{center}

\vspace*{.2cm}
\begin{center}
\begin{tabular}{|c|c|c|c|c|}
\hline
\hline
Baryon & Quark Content & $J^P$ & $(S_{qq}, I_{qq})$ & Mass (GeV)\\
\hline
\hline
$\Lambda_c^+$ &  c[ud] & ${\frac{1}{2}}^+$ & (0,0) & 2.285\\
\hline
$\Xi_c^+$ &  c[us] & ${\frac{1}{2}}^+$ & (0,1/2) & 2.470\\
\hline
$\Xi_c^0$ &  c[ds] & ${\frac{1}{2}}^+$ & (0,1/2) & 2.466\\
\hline
$\Xi_c^{\prime +}$ &  c\{us\} & ${\frac{1}{2}}^+$ & (1,1/2) & 2.470\\
\hline
$\Xi_c^{\prime 0}$ &  c\{ds\} & ${\frac{1}{2}}^+$ & (1,1/2) & 2.466\\
\hline
$\Sigma_c^0$ &  c\{dd\} & ${\frac{1}{2}}^+$ & (1,1) & 2.453\\
\hline
$\Omega_c^0$ &  c\{ss\} & ${\frac{1}{2}}^+$ & (1,0) & 2.704\\
\hline
$\Lambda_b^0$ &  b[ud] & ${\frac{1}{2}}^+$ & (0,0) & 5.640\\
\hline
$\Xi_b^0$ &  b[us] & ${\frac{1}{2}}^+$ & (0,1/2) & 5.800\\
\hline
$\Omega_b^-$ &  b\{ss\} & ${\frac{1}{2}}^+$ & (1,0) & 6.040\\
\hline
\end{tabular}
\end{center}

\newpage
\begin{center}
{\bf TABLE IIa}
\end{center}

\vspace*{.2cm}
\begin{center}
\begin{tabular}{|c|c|c|c|c|c|c|c|}
\hline\hline
Decay  & $\ell^-_{FD}$ & $\ell^+_{FD}$  & $\ell^{P_+}_{II_a}$
& $\ell^{Q_+}_{II_a}$  & $\ell^{D_+}_{II_b}$ & $\ell^{Q_+}_{II_b}$
& $\ell_{III}$  \\
\hline\hline
$\Lambda_b^0\to \Lambda_c^+\pi^-$& $-1$ & $-1$ & $-\frac{1}{2}$
& $\frac{1}{2}$ & $\frac{1}{2}$ & $\frac{1}{2}$ & $-2$ \\
\hline
$\Lambda_b^0\to \Sigma_c^+\pi^-$  & $0$ & $0$ & $\frac{\sqrt{3}}{2}$
& $\frac{1}{2\sqrt{3}}$ & $\frac{\sqrt{3}}{2}$ & $\frac{\sqrt{3}}{2}$
& $-2\sqrt{3}$ \\
\hline
$\Lambda_b^0\to \Sigma_c^0\pi^0$ & $0$ & $0$ & $-\frac{\sqrt{3}}{2}$
& $-\frac{1}{2\sqrt{3}}$ & $-\frac{\sqrt{3}}{2}$ & $-\frac{\sqrt{3}}{2}$
& $2\sqrt{3}$ \\
\hline
$\Lambda_b^0\to\Sigma_c^0\eta$ & $0$ & $0$
& $-\frac{\sqrt{3}}{2}S$ & $-\frac{1}{2\sqrt{3}}S$
& $\frac{\sqrt{3}}{2}S$ & $\frac{\sqrt{3}}{2}S$
& $2\sqrt{6}S$\\
\hline
$\Lambda_b^0\to\Sigma_c^0\eta^\prime$ & $0$ & $0$
& $\frac{\sqrt{3}}{2}C$ & $\frac{1}{2\sqrt{3}}C$
& $-\frac{\sqrt{3}}{2}C$ & $-\frac{\sqrt{3}}{2}C$
& $-2\sqrt{6}C$\\
\hline
$\Lambda_b^0\to \Xi_c^0 K^0$ & $0$ & $0$ & $-\frac{1}{2}$ &
$\frac{1}{2}$ & $0$ & $0$ & $-2$ \\
\hline
$\Lambda_b^0\to \Xi_c^{\prime 0} K^0$ & $0$ & $0$ & $\frac{\sqrt{3}}{2}$ &
$\frac{1}{2\sqrt{3}}$ & $0$ & $0$ & $-2\sqrt{3}$ \\
\hline
$\Xi_b^0\to \Xi_c^+\pi^-$  & $-1$ & $-1$ & $-\frac{1}{2}$
& $\frac{1}{2}$ & $0$ & $0$ & $0$ \\
\hline
$\Xi_b^0\to\Xi_c^{\prime +}\pi^-$ & $0$ & $0$
& $-\frac{\sqrt{3}}{2}$ & $-\frac{1}{2\sqrt{3}}$
& $0$ & $0$ & $0$\\
\hline
$\Xi_b^0\to \Xi_c^0\pi^0$ & $0$ & $0$ & $\frac{1}{2\sqrt{2}}$
& $-\frac{1}{2\sqrt{2}}$
& $\frac{1}{2\sqrt{2}}$ & $\frac{1}{2\sqrt{2}}$ & $0$ \\
\hline
$\Xi_b^0\to\Xi_c^0\eta$ & $0$ & $0$
& $\frac{1}{2\sqrt{2}}S$ & $-\frac{1}{2\sqrt{2}}S$
& $-\frac{1}{2\sqrt{2}}S$ & $-\frac{1}{2\sqrt{2}}S$
& $-2C$\\
\hline
$\Xi_b^0\to\Xi_c^0\eta^\prime$ & $0$ & $0$
& $-\frac{1}{2\sqrt{2}}C$ & $\frac{1}{2\sqrt{2}}C$
& $\frac{1}{2\sqrt{2}}C$ & $\frac{1}{2\sqrt{2}}C$
& $-2S$\\
\hline
$\Xi_b^0\to\Xi_c^{\prime 0}\pi^0$ & $0$ & $0$ & $\frac{\sqrt{3}}{2}$ &
$\frac{1}{2\sqrt{3}}$ & $\frac{\sqrt{3}}{2}$ & $\frac{\sqrt{3}}{2}$ & $0$ \\
\hline
$\Xi_b^0\to\Xi_c^{\prime 0}\eta$ & $0$ & $0$
& $\frac{\sqrt{3}}{2\sqrt{2}}S$
& $\frac{\sqrt{3}}{2\sqrt{2}}S$
& $-\frac{\sqrt{3}}{2\sqrt{2}}S$
& $-\frac{\sqrt{3}}{2\sqrt{2}}S$
& $-2\sqrt{3}C$\\
\hline
$\Xi_b^0\to\Xi_c^{\prime 0}\eta^\prime$ & $0$ & $0$
& $-\frac{\sqrt{3}}{2\sqrt{2}}C$
& $-\frac{\sqrt{3}}{2\sqrt{2}}C$
& $\frac{\sqrt{3}}{2\sqrt{2}}C$
& $\frac{\sqrt{3}}{2\sqrt{2}}C$
& $-2\sqrt{3}S$\\
\hline
$\Xi_b^0\to\Lambda_c^+ K^-$ & $0$ & $0$ & $0$ & $0$
& $-\frac{1}{2}$ & $-\frac{1}{2}$ & $2$ \\
\hline
$\Xi_b^0\to\Sigma_c^+ K^-$ & $0$ & $0$ & $0$ & $0$
& $-\frac{\sqrt{3}}{2}$ & $-\frac{\sqrt{3}}{2}$
& $2\sqrt{3}$\\
\hline
$\Xi_b^0\to\Sigma_c^0\bar K^0$ &$0$ &$0$ &$0$ &$0$ & $0$ & $0$ &$2\sqrt{6}$\\
\hline
$\Xi_b^0\to\Omega_c^0 K^0$ & $0$ & $0$ & $-\sqrt{\frac{3}{2}}$
& $-\frac{1}{\sqrt{6}}$ & $0$ & $0$ & $0$\\
\hline
$\Xi_b^-\to \Xi_c^0 \pi^-$ & $-1$ & $-1$
& $\frac{1}{2}$ & $\frac{1}{2}$ & $0$ & $0$ & $0$\\
\hline
$\Xi_b^-\to \Xi_c^{\prime 0}\pi^-$ & $0$ & $0$
& $0$ & $0$ & $\frac{\sqrt{3}}{2}$ & $\frac{\sqrt{3}}{2}$ & $0$ \\
\hline
$\Xi_b^-\to \Sigma_c^0 K^-$ & $0$ & $0$
& $0$ & $0$ & $-\sqrt{\frac{3}{2}}$ & $-\sqrt{\frac{3}{2}}$ & $0$ \\
\hline
$\Omega_b^-\to \Omega_c^0\pi^-$& $-1$ & $ \frac{1}{3}$
& $0$ & $0$& $0$ & $0$ & $0$\\
\hline\hline
\end{tabular}
\end{center}

\newpage
\begin{center}
{\bf TABLE IIb}
\end{center}

\vspace*{.2cm}
\begin{center}
\begin{tabular}{||l|c|c|c|c|c|c|c||}
\hline\hline
Decay  & $\ell^-_{FD}$ & $\ell^+_{FD}$  & $\ell^{P_+}_{II_a}$
& $\ell^{Q_+}_{II_a}$  & $\ell^{D_+}_{II_b}$ & $\ell^{Q_+}_{II_b}$
& $\ell_{III}$  \\
\hline\hline
$\Lambda_c^+\to \Lambda^0\pi^+$& $-1$ & $-1$ & $-\frac{1}{2}$
& $\frac{1}{2}$ & $\frac{1}{2}$ & $\frac{1}{2}$ & $-2$ \\
\hline
$\Lambda_c^+\to \Sigma^0\pi^+$  & $0$ & $0$ & $-\frac{\sqrt{3}}{2}$
& $-\frac{1}{2\sqrt{3}}$ & $-\frac{\sqrt{3}}{2}$ & $-\frac{\sqrt{3}}{2}$
& $2\sqrt{3}$ \\
\hline
$\Lambda_c^+\to \Sigma^+\pi^0$ & $0$ & $0$ & $\frac{\sqrt{3}}{2}$
& $\frac{1}{2\sqrt{3}}$ & $\frac{\sqrt{3}}{2}$ & $\frac{\sqrt{3}}{2}$
& $-2\sqrt{3}$ \\
\hline
$\Lambda_c^+\to \Sigma^+\eta$ & $0$ & $0$
& $-\frac{\sqrt{3}}{2} \cdot S \cdot {\rm ctg}_+$
& $-\frac{1}{2\sqrt{3}} \cdot S \cdot {\rm ctg}_-$
& $\frac{\sqrt{3}}{2}  \cdot S $ & $\frac{\sqrt{3}}{2} \cdot S $
& $\sqrt{3} \cdot S $ \\
\hline
$\Lambda_c^+\to \Sigma^+\eta^\prime$ & $0$ & $0$
& $\frac{\sqrt{3}}{2} \cdot C \cdot {\rm tg}_-$
& $\frac{1}{2\sqrt{3}} \cdot C \cdot {\rm tg}_+$
& $-\frac{\sqrt{3}}{2}  \cdot C $ & $-\frac{\sqrt{3}}{2} \cdot C $
& $-\sqrt{3} \cdot C $ \\
\hline
$\Lambda_c^+\to p\bar K^0$ & $\frac{3}{\sqrt{6}}$
& $\frac{3}{\sqrt{6}}$ & $-\frac{3}{\sqrt{6}}$
& $\frac{1}{\sqrt{6}}$ & $0$ & $0$ & $0$\\
\hline
$\Lambda_c^+\to \Xi^0 K^+$ & $0$ & $0$ & $0$ & $\frac{2}{\sqrt{6}}$
& $0$ & $0$ & $-2\sqrt{6}$ \\
\hline
$\Xi_c^+\to \Sigma^+\bar K^0$  & $\frac{3}{\sqrt{6}}$
& $\frac{3}{\sqrt{6}}$ & $0$ & $0$ & $\frac{3}{\sqrt{6}}$
& $\frac{3}{\sqrt{6}}$ & $0$ \\
\hline
$\Xi_c^+\to \Xi^0\pi^+$ & $\frac{3}{\sqrt{6}}$ & $\frac{3}{\sqrt{6}}$
& $0$ & $0$ & $-\frac{3}{\sqrt{6}}$ & $-\frac{3}{\sqrt{6}}$ & $0$ \\
\hline
$\Xi_c^0\to\Lambda^0\bar K^0$ & $-\frac{1}{2}$ & $-\frac{1}{2}$
& $1$ & $0$ & $\frac{1}{2}$ & $\frac{1}{2}$ & $-2$ \\
\hline
$\Xi_c^0\to\Sigma^0\bar K^0$  & $-\frac{\sqrt{3}}{2}$ &
$-\frac{\sqrt{3}}{2}$ & $0$ & $-\frac{1}{\sqrt{3}}$ & $-\frac{\sqrt{3}}{2}$
& $-\frac{\sqrt{3}}{2}$ & $2\sqrt{3}$ \\
\hline
$\Xi_c^0\to\Sigma^+ K^-$ & $0$ & $0$ & $0$ & $\frac{2}{\sqrt{6}}$ &
$0$ & $0$ & $-2\sqrt{6}$\\
\hline
$\Xi_c^0\to \Xi^0 \pi^0$ & $0$ & $0$ & $\frac{\sqrt{3}}{2}$
& $-\frac{1}{2\sqrt{3}}$ & $\frac{\sqrt{3}}{2}$ & $\frac{\sqrt{3}}{2}$ & $0$\\
\hline
$\Xi_c^0\to \Xi^0\eta$ & $0$ & $0$
& $-\frac{\sqrt{3}}{2} \cdot S \cdot {\rm ctg}_+$
& $\frac{1}{2\sqrt{3}} \cdot S \cdot {\rm ctg}_-$
& $\frac{\sqrt{3}}{2}  \cdot S $ & $\frac{\sqrt{3}}{2} \cdot S $
& $\sqrt{6} \cdot C$ \\
\hline
$\Xi_c^0\to \Xi^0\eta^\prime$ & $0$ & $0$
& $\frac{\sqrt{3}}{2} \cdot C \cdot {\rm tg}_-$
& $-\frac{1}{2\sqrt{3}} \cdot C \cdot {\rm tg}_+$
& $-\frac{\sqrt{3}}{2}  \cdot C $ & $-\frac{\sqrt{3}}{2} \cdot C $
& $\sqrt{6} \cdot S $ \\
\hline
$\Xi_c^0\to \Xi^- \pi^+$ & $-\frac{3}{\sqrt{6}}$ & $-\frac{3}{\sqrt{6}}$
& $-\frac{3}{\sqrt{6}}$ & $\frac{1}{\sqrt{6}}$ & $0$ & $0$ & $0$\\
\hline
$\Omega_c^0\to \Xi^0\bar K^0$ & $-1$ & $\frac{1}{3}$
& $0$ & $0$ & $-3$ & $1$ & $0$ \\
\hline
$\Lambda_b^0\to \Lambda^0\pi^0$ & $-\frac{1}{\sqrt{2}}$
& $-\frac{1}{\sqrt{2}}$ & $-\frac{1}{2\sqrt{2}}$ & $\frac{1}{2\sqrt{2}}$
& $\frac{1}{2\sqrt{2}}$ & $\frac{1}{2\sqrt{2}}$  & $\sqrt{2}$ \\
\hline
$\Lambda_b^0\to p K^-$& $\frac{3}{\sqrt{6}}$ & $\frac{3}{\sqrt{6}}$
& $\frac{3}{\sqrt{6}}$ & $-\frac{1}{\sqrt{6}}$ & $0$ & $0$ & $0$ \\
\hline\hline
\end{tabular}
\end{center}

\newpage
\begin{center}
{\bf TABLE IIIa}
\end{center}

\vspace*{.2cm}
\begin{center}
\begin{tabular}{||c|c|c||}
\hline\hline
Decay & $t_2(r)$ & $t_3(r)$\\
\hline\hline
$\Lambda_b^0\to \Lambda_c^+\pi^-$ & $1$ & $1$\\
\hline
$\Lambda_b^0\to \Sigma_c^+\pi^-$ & $1$ & $1$\\
\hline
$\Lambda_b^0\to \Sigma_c^0\pi^0$ & $1$ & $1$\\
\hline
$\Lambda_b^0\to \Sigma_c^0\eta$
& $r^2/\sqrt{C^2 \cdot r^2 + S^2}$ & $r^2/\sqrt{C^2 \cdot r^2 + S^2}$\\
\hline
$\Lambda_b^0\to \Sigma_c^0\eta^\prime$
& $r^2/\sqrt{S^2 \cdot r^2 + C^2}$ & $r^2/\sqrt{S^2 \cdot r^2 + C^2}$\\
\hline
$\Lambda_b^0\to \Xi_c^0 K^0$ & $(1+r)^2/4$ & $(1+r)^2/4$\\
\hline
$\Lambda_b^0\to \Xi_c^{\prime 0} K^0$ & $(1+r)^2/4$ & $(1+r)^2/4$\\
\hline
$\Xi_b^0\to \Xi_c^+ \pi^-$ & $1$ & $1$\\
\hline
$\Xi_b^0\to \Xi_c^{\prime +} \pi^-$ & $1$ & $1$\\
\hline
$\Xi_b^0\to \Xi_c^0 \pi^0$ & $1$ & $1$\\
\hline
$\Xi_b^0\to \Xi_c^0 \eta$ & $r^2/\sqrt{C^2 \cdot r^2 + S^2}$
& $r^3/\sqrt{C^2 \cdot r^2 + S^2}$\\
\hline
$\Xi_b^0\to \Xi_c^0 \eta^\prime$ & $r^2/\sqrt{S^2 \cdot r^2 + C^2}$
& $r^3/\sqrt{S^2 \cdot r^2 + C^2}$\\
\hline
$\Xi_b^0\to \Xi_c^{\prime 0} \pi^0$ & $1$ & $1$\\
\hline
$\Xi_b^0\to \Xi_c^{\prime 0} \eta$ & $r^2/\sqrt{C^2 \cdot r^2 + S^2}$
& $r^3/\sqrt{C^2 \cdot r^2 + S^2}$\\
\hline
$\Xi_b^0\to \Xi_c^{\prime 0} \eta^\prime$ & $r^2/\sqrt{S^2 \cdot r^2 + C^2}$
& $r^3/\sqrt{S^2 \cdot r^2 + C^2}$\\
\hline
$\Xi_b^0\to \Lambda_c^+ K^-$ & $(1+r)^2/4$ & $(1+r)^2/4$\\
\hline
$\Xi_b^0\to \Sigma_c^+ K^-$ & $(1+r)^2/4$ & $(1+r)^2/4$\\
\hline
$\Xi_b^0\to \Sigma_c^0 \bar K^0$ & $0$ & $(1+r)^2/4$\\
\hline
$\Xi_b^0\to \Omega_c^0 K^0$ & $(1+r)^2/4$ & $0$\\
\hline
$\Xi_b^-\to \Xi_c^0 \pi^-$ & $1$ & $1$ \\
\hline
$\Xi_b^-\to \Xi_c^{\prime 0}\pi^-$ & $1$ & $1$ \\
\hline
$\Xi_b^-\to \Sigma_c^0 K^-$ & $(1+r)^2/4$ & $0$\\
\hline
$\Omega_b^-\to \Omega_c^0\pi^-$ & $1$ & $1$ \\
\hline\hline
\end{tabular}
\end{center}

\newpage
\begin{center}
{\bf TABLE IIIb}
\end{center}

\vspace*{.2cm}
\begin{center}
\begin{tabular}{||c|c|c|c||}
\hline\hline
Decay & $\chi_f$ & $t_2(r)$ & $t_3(r)$\\
\hline\hline
$\Lambda_c^+\to \Lambda^0\pi^+$ & $\frac{2}{3r}+\frac{1}{3}$ & $1$ & $1$\\
\hline
$\Lambda_c^+\to \Sigma^0\pi^+$ & $\frac{2}{3r}+\frac{1}{3}$ & $1$ & $1$\\
\hline
$\Lambda_c^+\to \Sigma^+\pi^0$ & $\frac{2}{3r}+\frac{1}{3}$ & $1$ & $1$\\
\hline
$\Lambda_c^+\to \Sigma^+\eta$ & $\frac{2}{3r}+\frac{1}{3}$  & $1$ & $1$\\
\hline
$\Lambda_c^+\to \Sigma^+\eta^\prime$&$\frac{2}{3r}+\frac{1}{3}$ & $1$ & $1$\\
\hline
$\Lambda_c^+\to p\bar K^0$ &$1$ &$(1+r)^2\sqrt{r}/4$ & $(1+r)^2\sqrt{r}/4$\\
\hline
$\Lambda_c^+\to \Xi^0 K^+$ & $\frac{2r}{3}+\frac{1}{3}$
& $(1+r)^2\sqrt{r}/4$ & $(1+r)^2\sqrt{r}/4$\\
\hline
$\Xi_c^+\to \Sigma^+\bar K^0$ &$\frac{2}{3} + \frac{r}{3}$
& $(1+r)^2\sqrt{r}/4$ & $(1+r)^2\sqrt{r}/4$\\
\hline
$\Xi_c^+\to \Xi^0 \pi^+$ & $\frac{2}{3} + \frac{1}{3r}$ & $1$ & $1$\\
\hline
$\Xi_c^0\to \Lambda^0\bar K^0$ & $\frac{2}{3} + \frac{r}{3}$
& $(1+r)^2\sqrt{r}/4$ & $(1+r)^2\sqrt{r}/4$\\
\hline
$\Xi_c^0\to \Sigma^0\bar K^0$ & $\frac{2}{3} + \frac{r}{3}$
& $(1+r)^2\sqrt{r}/4$ & $(1+r)^2\sqrt{r}/4$\\
\hline
$\Xi_c^0\to \Sigma^+ K^-$ & $\frac{2}{3}  + \frac{r}{3}$
& $(1+r)^2\sqrt{r}/4$ & $(1+r)^2\sqrt{r}/4$\\
\hline
$\Xi_c^0\to \Xi^0 \pi^0$  & $\frac{2}{3} + \frac{1}{3r}$ & $1$ & $1$\\
\hline
$\Xi_c^0\to \Xi^0\eta$  & $\frac{2}{3} + \frac{1}{3r}$ & $1$ & $r$\\
\hline
$\Xi_c^0\to \Xi^0\eta^\prime$ & $\frac{2}{3} + \frac{1}{3r}$ & $1$ & $r$\\
\hline
$\Xi_c^0\to \Xi^- \pi^+$  & $\frac{2}{3} + \frac{1}{3r}$ & $1$ & $1$\\
\hline
$\Omega_c^0\to \Xi^0\bar K^0$ & $\frac{2r}{3} + \frac{1}{3}$
& $(1+r)^2\sqrt{r}/4$ & $(1+r)^2\sqrt{r}/4$\\
\hline
$\Lambda_b^0\to \Lambda^0\pi^0$ & $\frac{2}{3r} + \frac{1}{3}$ & $1$ & $1$\\
\hline
$\Lambda_b^0\to p K^-$ & $1$ & $(1+r)^2\sqrt{r}/4$ & $(1+r)^2\sqrt{r}/4$\\
\hline\hline
\end{tabular}
\end{center}

\newpage
\begin{center}
{\bf TABLE IV}
\end{center}

\vspace*{.2cm}
\begin{center}
\begin{tabular}{|c|c|c|c|c|c|}
\hline\hline
Process & K\"{o}rner,             & Xu,
& Cheng,             & \,\,\,\,\, Our \,\,\,\,\, & Experiment \cite{PDG}\\
        & Kr\"{a}mer \cite{Kramer}& Kamal \cite{Kamal}
& Tseng \cite{Cheng1} & & \\
\hline
$\Lambda^+_c\to \Lambda \pi^+$ & 0.76 & 1.67 & 0.91 & 0.79 & 0.79$\pm$ 0.18\\
\hline
$\Lambda^+_c\to \Sigma^0 \pi^+$ & 0.33 & 0.35 & 0.74 & 0.88 & 0.88$\pm$ 0.20\\
\hline
$\Lambda^+_c\to \Sigma^+ \pi^0$ & 0.33 & 0.35 & 0.74 & 0.88 & 0.88$\pm$ 0.22\\
\hline
$\Lambda^+_c\to \Sigma^+ \eta$ & 0.16 & & & 0.11 & 0.48$\pm$ 0.17\\
\hline
$\Lambda^+_c\to \Sigma^+ \eta^\prime$ & 1.28 & & & 0.12& \\
\hline
$\Lambda^+_c\to p \bar K^0$ & 2.16 & 1.24 & 1.30 & 2.06 & 2.2$\pm$ 0.4\\
\hline
$\Lambda^+_c\to \Xi^0 K^+$ & 0.27 & 0.10 & & 0.31 &  0.34$\pm$ 0.09\\
\hline
$\Xi^+_c\to \Sigma^+ \bar K^0$ & 5.11 & 0.35 & 0.67 & 3.08 & \\
\hline
$\Xi^+_c\to \Xi^0 \pi^+$ & 2.80 & 2.66 & 3.12 & 4.40 & 1.2$\pm$0.5$\pm$0.3\\
\hline
$\Xi^0_c\to \Lambda \bar K^0$ & 0.11 & 0.32 & 0.24 & 0.42 & \\
\hline
$\Xi^0_c\to \Sigma^0 \bar K^0$ & 1.03 & 0.08 & 0.12 & 0.20 & \\
\hline
$\Xi^0_c\to \Sigma^+ K^-$ & 0.11 & 0.11 &  & 0.27 & \\
\hline
$\Xi^0_c\to \Xi^0 \pi^0$ & 0.03 & 0.49 & 0.25 & 0.04 & \\
\hline
$\Xi^0_c\to \Xi^0 \eta$ & 0.21 & & & 0.28 & \\
\hline
$\Xi^0_c\to \Xi^0 \eta^\prime$ & 0.74 & & & 0.31 & \\
\hline
$\Xi^0_c\to \Xi^- \pi^+$ & 0.91 & 1.52 & 1.10 & 1.22 & \\
\hline
$\Omega^0_c\to \Xi^0 \bar K^0$ & 1.10 & & 0.08 & 0.02 & \\
\hline
$\Lambda^0_b\to \Lambda \pi^0$ & & & & 4.92$\times$10$^{-5}$ & \\
\hline
$\Lambda^0_b\to p K^-$ & & & & 2.11$\times$10$^{-4}$ & \\
\hline\hline
\end{tabular}
\end{center}

\newpage
\vspace*{.5cm}
\begin{center}
{\bf TABLE V}
\end{center}

\vspace*{.2cm}
\begin{center}
\begin{tabular}{|c|c|c|c|c|c|}
\hline\hline
Process & K\"{o}rner, & Xu,
        & Cheng,      & \,\,\,\,\, Our \,\,\,\,\, & Experiment \cite{PDG}\\
        & Kr\"{a}mer \cite{Kramer}& Kamal \cite{Kamal}
        & Tseng \cite{Cheng1} & & \\
\hline
$\Lambda^+_c\to \Lambda \pi^+$ & -0.70 & -0.67 & -0.95 & -0.95 & -0.98 $\pm$ 0.19\\
\hline
$\Lambda^+_c\to \Sigma^0 \pi^+$ & 0.70 & 0.92 & 0.78 & 0.43 & \\
\hline
$\Lambda^+_c\to \Sigma^+ \pi^0$ & 0.71 & 0.92 & 0.78 & 0.43 &
 -0.45$\pm$ 0.31$\pm$ 0.06 \\
\hline
$\Lambda^+_c\to \Sigma^+ \eta$ & 0.33 & & & 0.55 & \\
\hline
$\Lambda^+_c\to \Sigma^+ \eta^\prime$ & -0.45 & & & -0.05 & \\
\hline
$\Lambda^+_c\to p \bar K^0$ & -1.0 & 0.51 & -0.49 & -0.97 & \\
\hline
$\Lambda^+_c\to \Xi^0 K^+$ & 0 & 0 & & 0 & \\
\hline
$\Xi^+_c\to \Sigma^+ \bar K^0$ & -1.0 & 0.24 & -0.09 & -0.99 & \\
\hline
$\Xi^+_c\to \Xi^0 \pi^+$ & -0.78 & -0.81 & -0.77 & -1.0 & \\
\hline
$\Xi^0_c\to \Lambda \bar K^0$ & -0.76 & 1.0 & -0.73 & -0.75 & \\
\hline
$\Xi^0_c\to \Sigma^0 \bar K^0$ & -0.96 & -0.99 & -0.59 & -0.55 & \\
\hline
$\Xi^0_c\to \Sigma^+ K^-$ & 0 & 0 &  & 0 & \\
\hline
$\Xi^0_c\to \Xi^0 \pi^0$ & 0.92 & 0.92 & -0.54 & 0.94 & \\
\hline
$\Xi^0_c\to \Xi^0 \eta$ & -0.92 & & & -1.0 & \\
\hline
$\Xi^0_c\to \Xi^0 \eta^\prime$ & -0.38 & & & -0.32 & \\
\hline
$\Xi^0_c\to \Xi^- \pi^+$ & -0.38 & -0.38 & -0.99 & -0.84& \\
\hline
$\Omega^0_c\to \Xi^0 \bar K^0$   & 0.51 &  & -0.93 & -0.81 & \\
\hline
$\Lambda^0_b\to \Lambda \pi^0$ & & & & -1.0 &\\
\hline
$\Lambda^0_b\to p K^-$ & & & & -0.88 & \\
\hline\hline
\end{tabular}
\end{center}

\newpage
\begin{center}
{\bf TABLE VI}
\end{center}

\vspace*{.2cm}
\begin{center}
\begin{tabular}{|c||c|c|c|c|}
\hline\hline
Reference     & \multicolumn{2}{|c|} {$\Lambda_c^+\to\Lambda\pi^+$}
              & \multicolumn{2}{|c|} {$\Lambda_c^+\to\Sigma^+\pi^0$}
 \\
\cline{2-5}  & $A$  & $B$ & $A$ & $B$ \\
\hline\hline
CLEO II \cite{CLEO2} & -3.0$^{+0.8}_{-1.2}$ & 12.7$^{+2.7}_{-2.5}$
& 1.3$^{+0.9}_{-1.1}$ & -17.3$^{+2.3}_{-2.9}$\\
\hline
Xu and Kamal \cite{Kamal}& -2.7 & 20.8 & -2.9 &  -6.0  \\
\hline
Cheng and Tseng \cite{Cheng} & -3.5 & 13.2 & -2.4 &  -14.6 \\
\hline
K\"{o}rner and Kr\"{a}mer \cite{Kramer} & -1.9 & 13.9 & -1.3 &  -9.9 \\
\hline
Our & -4.2 & 9.0 & -1.2 &  -17.2 \\
\hline\hline
\end{tabular}
\end{center}

\vspace*{1cm}
\begin{center}
{\bf TABLE VII}
\end{center}

\vspace*{.2cm}
\begin{center}
\begin{tabular}{|c|c|c||c|c|c|}
\hline\hline
Process  & $\Gamma$ (in 10$^{10}$ s$^{-1}$) & $\alpha$  &
Process  & $\Gamma$ (in 10$^{10}$ s$^{-1}$) & $\alpha$  \\
\hline\hline
$\Lambda_b^0\to \Lambda_c^+\pi^-$     & 0.382 & -0.99 &
$\Xi_b^0\to\Xi_c^{\prime 0}\pi^0$ & 0.014 & 0.94\\
\hline
$\Lambda_b^0\to \Sigma_c^+\pi^-$      & 0.039 & 0.65 &
$\Xi_b^0\to\Xi_c^{\prime 0}\eta$ & 0.015 & -0.98\\
\hline
$\Lambda_b^0\to \Sigma_c^0\pi^0$      & 0.039 & 0.65 &
$\Xi_b^0\to\Xi_c^{\prime 0}\eta^\prime$& 0.021 & 0.97\\
\hline
$\Lambda_b^0\to\Sigma_c^0\eta$        & 0.023 & 0.79 &
$\Xi_b^0\to\Lambda_c^+ K^-$ & 0.010 & -0.73\\
\hline
$\Lambda_b^0\to\Sigma_c^0\eta^\prime$ & 0.029 & 0.99&
$\Xi_b^0\to\Sigma_c^+ K^-$ & 0.030 & -0.74\\
\hline
$\Lambda_b^0\to \Xi_c^0 K^0$          & 0.021 & -0.81 &
$\Xi_b^0\to\Sigma_c^0\bar K^0$ & 0.021 & 0\\
\hline
$\Lambda_b^0\to \Xi_c^{\prime 0} K^0$ & 0.032 & 0.98 &
$\Xi_b^0\to\Omega_c^0 K^0$ & 0.023 & 0.65\\
\hline
$\Xi_b^0\to \Xi_c^+\pi^-$ & 0.479 & -1.00 &
$\Xi_b^-\to \Xi_c^0 \pi^-$ & 0.645 & -0.97\\
\hline
$\Xi_b^0\to\Xi_c^{\prime +}\pi^-$& 0.018 & 0.61 &
$\Xi_b^-\to \Xi_c^{\prime 0}\pi^-$& 0.007 &-1.00\\
\hline
$\Xi_b^0\to \Xi_c^0\pi^0$ & 0.002 & -0.99 &
$\Xi_b^-\to \Sigma_c^0 K^-$ & 0.016 & -0.98\\
\hline
$\Xi_b^0\to\Xi_c^0\eta$ & 0.012 & -0.86 &
$\Omega_b^-\to \Omega_c^0\pi^-$ & 0.352 & 0.60\\
\hline
$\Xi_b^0\to\Xi_c^0\eta^\prime$ & 0.003 & 0.71 & & &\\
\hline\hline
\end{tabular}
\end{center}

\newpage
\begin{center}
{\bf TABLE VIII}
\end{center}

\vspace*{.2cm}
\begin{center}
\begin{tabular}{|c|c|c|c|c|}
\hline\hline
Amplitude   & \multicolumn{4}{|c|} {Diagram} \\
\cline{2-5}  & $II_a$  & $II_b$ & $II_a+II_b$ & $III$ \\
\hline\hline
A   & -29.8$\%$ & -18.5$\%$ & -48.3$\%$ & \\
\hline
B   & -32.4$\%$ & -15.9$\%$ & -48.3$\%$ &  -13.9$\%$  \\
\hline\hline
\end{tabular}
\end{center}

\vspace*{1cm}
\begin{center}
{\bf TABLE IX}
\end{center}

\vspace*{.2cm}
\begin{center}
\begin{tabular}{|c|c|c|c|c|}
\hline\hline
Amplitude   & \multicolumn{4}{|c|} {Diagram} \\
\cline{2-5}  & $II_a$  & $II_b$ & $II_a+II_b$ & $III$ \\
\hline\hline
A   & -13.9$\%$ & -6.2$\%$ & -20.1$\%$ & \\
\hline
B   & -14.3$\%$ & -5.8$\%$ & -20.1$\%$ &  -8.5$\%$  \\
\hline\hline
\end{tabular}
\end{center}

\newpage
\begin{center}
{\bf TABLE X}
\end{center}

\vspace*{.2cm}
\begin{center}
\begin{tabular}{|c|c|}
\hline\hline
Ratio of interest & {\rm Br}(p$\phi$)/{\rm Br}(pK$^-\pi^+$) (in $\%$)\\
\hline\hline
CLEO \cite{CLEO} & 0.024$\pm$0.006$\pm$0.003 \\
\hline
NA32 \cite{NA32} & 0.04$\pm$0.03 \\
\hline
K\"{o}rner \& Kr\"{a}mer \cite{Kramer} & 0.05 \\
\hline
Cheng \& Tseng \cite{Cheng2} &  0.016 \\
\hline
Datta \cite{Datta} &  0.01 \\
\hline
     & 0.022 ($a^\star_2$=-0.30)\\
     & 0.030 ($a^\star_2$=-0.35)\\
     & 0.040 ($a^\star_2$=-0.40)\\
Our  & 0.050 ($a^\star_2$=-0.45)\\
     & 0.062 ($a^\star_2$=-0.50)\\
     & 0.075 ($a^\star_2$=-0.55)\\
     & 0.090 ($a^\star_2$=-0.60)\\
     & 0.105 ($a^\star_2$=-0.65)\\
\hline\hline
\end{tabular}
\end{center}

\newpage
\begin{center}
{\bf TABLE XI}
\end{center}

\vspace*{.2cm}
\begin{center}
\begin{tabular}{|c|c|}
\hline\hline
Ratio of interest & {\rm Br}$(\Lambda_b^0\to J/\psi\Lambda)$ (in $\%$)\\
\hline\hline
UA1 \cite{PDG,UA1}& 1.4$\pm$0.9 \\
\hline
OPAL \cite{OPAL}& $<$ 1.1 \\
\hline
CDF \cite{CDF}& 0.037$\pm$0.017$\pm$0.004 \\
\hline
Cheng \& Tseng \cite{Cheng2} &  0.011 \\
\hline
Cheng \cite{Cheng3} &  0.016 \\
\hline
    & 0.027 ($a_2=$0.10) \\
    & 0.061 ($a_2=$0.15) \\
Our & 0.108 ($a_2=$0.20) \\
    & 0.169 ($a_2=$0.25) \\
    & 0.243 ($a_2=$0.25) \\
\hline\hline
\end{tabular}
\end{center}

\vspace*{1cm}
\begin{center}
{\bf TABLE XII}
\end{center}

\vspace*{.2cm}
\begin{center}
\begin{tabular}{|c|c|}
\hline
\hline
Baryon & Lagrangian\\
\hline
\hline
$p$ &   $g_p\bar p\gamma^\mu\gamma^5 V_+ d^{a_1}u^{a_2}C\gamma_\mu
V_+ u^{a_3} \varepsilon^{a_1a_2a_3} =
2 g_p\bar p V_+ u^{a_1}u^{a_2}C\gamma_5 V_+ d^{a_3}
\varepsilon^{a_1a_2a_3}$ \\
\hline
$n$ &   $ -g_n\bar n\gamma^\mu\gamma^5 V_+ u^{a_1}d^{a_2}C\gamma_\mu
V_+ d^{a_3} \varepsilon^{a_1a_2a_3} = - 2 g_n\bar n V_+ d^{a_1}d^{a_2}
C\gamma_5 V_+ u^{a_3} \varepsilon^{a_1a_2a_3}$ \\
\hline
$\Sigma^+$ &  $ - g_{\Sigma^+}\bar\Sigma^+ \gamma^\mu\gamma^5
V_+ s^{a_1}u^{a_2}C\gamma_\mu V_+ u^{a_3}
\varepsilon^{a_1a_2a_3} = - 2 g_{\Sigma^+}\bar\Sigma^+V_+
u^{a_1}u^{a_2}C\gamma_5 V_+ s^{a_3} \varepsilon^{a_1a_2a_3}$ \\
\hline
$\Sigma^-$ &  $g_{\Sigma^-}\bar\Sigma^-\gamma^\mu\gamma^5 V_+ s^{a_1}d^{a_2}
C\gamma_\mu V_+ d^{a_3} \varepsilon^{a_1a_2a_3} =
2 g_{\Sigma^-} \bar\Sigma^- V_+ d^{a_1}d^{a_2}
C\gamma_5 V_+ s^{a_3} \varepsilon^{a_1a_2a_3}$ \\
\hline
$\Sigma^0$ & $\sqrt{\frac{1}{2}}g_{\Sigma^0}\bar\Sigma^0\gamma^\mu\gamma^5
V_+ s^{a_1} (u^{a_2}C\gamma_\mu V_+ d^{a_3}  + d^{a_2}C\gamma_\mu V_+ u^{a_3})
\varepsilon^{a_1a_2a_3} = $ \\
 & $ = \sqrt{2} g_{\Sigma^0}\bar\Sigma^0 V_+ ( d^{a_1}u^{a_2}C\gamma_5
V_+ s^{a_3} + u^{a_1}d^{a_2}C\gamma_5 V_+ s^{a_3})\varepsilon^{a_1a_2a_3}$ \\
\hline
$\Lambda^0$ & $\sqrt{\frac{2}{3}}g_{\Lambda^0}\bar\Lambda^0\gamma^\mu\gamma^5
V_+ ( u^{a_1} d^{a_2}C\gamma_\mu V_+ s^{a_3}  - d^{a_1} u^{a_2}C\gamma_\mu
V_+ s^{a_3}) \varepsilon^{a_1a_2a_3} = $ \\
& $ = - \sqrt{6} g_{\Lambda^0}\bar\Lambda^0
V_+ s^{a_1}u^{a_2}C\gamma_5 V_+ d^{a_3}\varepsilon^{a_1a_2a_3}$\\
\hline
$\Xi^0$ & $g_{\Xi^0}\bar\Xi^0\gamma^\mu\gamma^5 V_+ u^{a_1}s^{a_2}C\gamma_\mu
V_+ s^{a_3} \varepsilon^{a_1a_2a_3} = 2 g_{\Xi^0}\bar\Xi^0 V_+ s^{a_1}s^{a_2}
C\gamma_5 V_+ u^{a_3} \varepsilon^{a_1a_2a_3}$ \\
\hline
$\Xi^-$ & $g_{\Xi^-}\bar\Xi^-\gamma^\mu\gamma^5 V_+ d^{a_1}s^{a_2}C\gamma_\mu
V_+ s^{a_3} \varepsilon^{a_1a_2a_3} = 2 g_{\Xi^-}\bar\Xi^-
V_+ s^{a_1}s^{a_2}C\gamma_5 V_+ d^{a_3} \varepsilon^{a_1a_2a_3}$ \\
\hline\hline
\end{tabular}
\end{center}

\newpage
\begin{center}
{\bf TABLE XIII}
\end{center}

\vspace*{.2cm}
\begin{center}
\begin{tabular}{|c|c|}
\hline
\hline
Baryon & Lagrangian \\
\hline
\hline
$\Lambda_Q$ &  - $g_{\Lambda_Q}\bar\Lambda_Q
Q^{a_1}u^{a_2}C\gamma_5 V_+ d^{a_3} \varepsilon^{a_1a_2a_3}$ \\
\hline
$\Xi_Q$ & $g_{\Xi_Q}\bar\Xi_Q
Q^{a_1}u^{a_2}C\gamma_5 V_+ s^{a_3}\varepsilon^{a_1a_2a_3}$\\
        & $g_{\Xi_Q}\bar\Xi_Q
Q^{a_1}d^{a_2}C\gamma_5 V_+ s^{a_3}\varepsilon^{a_1a_2a_3}$\\
\hline
          & $\frac{1}{\sqrt{6}}g_{\Sigma_Q}\bar\Sigma_Q \gamma^\mu\gamma^5
Q^{a_1}u^{a_2}C\gamma_\mu V_+ u^{a_3}\varepsilon^{a_1a_2a_3}$\\
$\Sigma_Q$& $\frac{1}{\sqrt{6}}g_{\Sigma_Q}\bar\Sigma_Q \gamma^\mu\gamma^5
Q^{a_1}d^{a_2}C\gamma_\mu V_+ d^{a_3}\varepsilon^{a_1a_2a_3}$\\
         &$\frac{1}{\sqrt{3}}g_{\Sigma_Q}\bar\Sigma_Q \gamma^\mu\gamma^5
Q^{a_1}u^{a_2}C\gamma_\mu V_+ d^{a_3}\varepsilon^{a_1a_2a_3}$\\
\hline
$\Xi_Q^\prime$ & $g_{\Xi_Q^\prime}\bar\Xi_Q^\prime \gamma^\mu\gamma^5
Q^{a_1}u^{a_2}C\gamma_\mu V_+ s^{a_3}\varepsilon^{a_1a_2a_3}$\\
        & $g_{\Xi_Q^\prime}\bar\Xi_Q^\prime \gamma^\mu\gamma^5
Q^{a_1}d^{a_2}C\gamma_\mu V_+ s^{a_3}\varepsilon^{a_1a_2a_3}$\\
\hline
$\Omega_Q$&
$\frac{1}{\sqrt{6}}g_{\Omega_Q}\bar\Omega_Q \gamma^\mu\gamma^5
Q^{a_1}s^{a_2}C\gamma_\mu V_+ s^{a_3}\varepsilon^{a_1a_2a_3}$ \\
\hline\hline
\end{tabular}
\end{center}

\vspace*{1cm}
\begin{center}
{\bf TABLE XIV}
\end{center}

\vspace*{.2cm}
\begin{center}
\begin{tabular}{|c|c|}
\hline
\hline
Meson & Lagrangian \\
\hline
\hline
$\pi$ & $g_\pi \pi^+ \bar u i\gamma^5  d + h.c.$ \\
\hline
$K$ & $g_K \left[ K^+ \bar u  i\gamma^5 V_+  s
+ K^0 \bar d i\gamma^5 V_+ s \right] + $ h.c.\\
\hline
$\phi$ & $g_\phi \phi_\mu \bar s  (\gamma^\mu - v^\mu) V_+  s$\\
\hline
$J/\psi$ & $g_{J/\psi}  (J/\psi)_\mu \bar c (\gamma^\mu - v^\mu) V_+  c$\\
\hline\hline
\end{tabular}
\end{center}

\newpage

\unitlength=1.00mm
\special{em:linewidth 0.4pt}
\linethickness{0.4pt}
\begin{picture}(127.00,116.00)
\put(70.00,23.00){\oval(48.00,16.00)[]}
\put(45.00,23.00){\circle*{5.00}}
\put(95.00,23.00){\circle*{5.00}}
\put(95.00,23.00){\line(1,0){25.00}}
\put(120.00,24.00){\line(-1,0){25.00}}
\put(95.00,22.00){\line(1,0){25.00}}
\put(45.00,22.00){\line(-1,0){25.00}}
\put(20.00,23.00){\line(1,0){25.00}}
\put(45.00,24.00){\line(-1,0){25.00}}
\put(44.00,24.00){\line(3,5){26.00}}
\put(96.00,24.00){\line(-3,5){26.00}}
\put(70.00,67.00){\circle*{2.00}}
\put(75.00,92.00){\line(0,3){8.00}}
\put(75.00,100.00){\line(-1,-3){1.50}}
\put(75.00,100.00){\line(1,-3){1.50}}
\put(90.00,96.00){\makebox(0,0)[cc]{{\bf p$_3$=p$_1$-p$_2$}}}
\put(10.00,23.00){\makebox(0,0)[cc]{{\large\bf B$_{Q^\prime (q)}$}}}
\put(128.00,23.00){\makebox(0,0)[cc]{{\large\bf B$_Q$}}}
\put(110.00,29.00){\makebox(0,0)[cc]{{\bf p$_1$}}}
\put(30.00,29.00){\makebox(0,0)[cc]{{\bf p$_2$}}}
\put(94.00,47.00){\makebox(0,0)[cc]{{\bf k+p$_1$}}}
\put(47.00,47.00){\makebox(0,0)[cc]{{\bf k+p$_2$}}}
\put(71.00,36.00){\makebox(0,0)[cc]{{\bf k$^\prime$}}}
\put(68.00,10.00){\makebox(0,0)[cc]{{\bf - (k$^\prime$+k)}}}
\put(28.00,23.00){\line(2,1){4.00}}
\put(28.00,23.00){\line(2,-1){4.00}}
\put(108.00,23.00){\line(2,1){4.00}}
\put(108.00,23.00){\line(2,-1){4.00}}
\put(68.00,31.10){\line(4,1){4.00}}
\put(68.00,31.10){\line(4,-1){4.00}}
\put(68.00,15.10){\line(4,1){4.00}}
\put(68.00,15.10){\line(4,-1){4.00}}
\put(56.00,44.00){\line(0,2){4.00}}
\put(56.00,44.00){\line(2,1){4.00}}
\put(82.00,47.30){\line(2,-1){4.00}}
\put(82.00,47.30){\line(0,-2){4.00}}
\put(70.00,78.00){\circle{14.00}}
\put(70.00,85.00){\circle*{2.00}}
\put(70.00,71.00){\circle*{2.00}}
\put(70.50,85.00){\line(0,1){15.00}}
\put(69.50,85.00){\line(0,1){15.00}}
\put(70.00,104.00){\makebox(0,0)[cc]{{\large\bf M}}}
\put(70.00,74.00){\makebox(0,0)[cc]{{\bf O$_\mu$}}}
\put(75.00,86.50){\makebox(0,0)[cc]{{\bf $\Gamma_M$}}}
\put(76.00,66.00){\makebox(0,0)[cc]{{\bf O$_\mu$}}}
\end{picture}
\begin{center}
\vspace*{1cm}
\hspace*{-1cm}Diagram I
\end{center}

\newpage
\unitlength=1.00mm
\special{em:linewidth 0.4pt}
\linethickness{0.4pt}
\begin{picture}(127.00,116.00)
\put(70.00,23.00){\oval(48.00,16.00)[]}
\put(46.00,23.00){\circle*{5.00}}
\put(94.00,23.00){\circle*{5.00}}
\put(95.00,23.00){\line(1,0){25.00}}
\put(120.00,24.00){\line(-1,0){25.00}}
\put(95.00,22.00){\line(1,0){25.00}}
\put(45.00,23.00){\line(1,0){50.00}}
\put(45.00,22.00){\line(-1,0){25.00}}
\put(20.00,23.00){\line(1,0){25.00}}
\put(45.00,24.00){\line(-1,0){25.00}}
\put(65.00,02.00){\line(0,3){8.00}}
\put(65.00,02.00){\line(-1,3){1.50}}
\put(65.00,02.00){\line(1,3){1.50}}
\put(70.00,05.00){\makebox(0,0)[cc]{{\bf p$_3$}}}
\put(13.00,23.00){\makebox(0,0)[cc]{{\large\bf B$_{Q^\prime(q)}$}}}
\put(127.00,23.00){\makebox(0,0)[cc]{{\large\bf B$_{Q}$}}}
\put(110.00,29.00){\makebox(0,0)[cc]{{\bf p$_1$}}}
\put(30.00,29.00){\makebox(0,0)[cc]{{\bf p$_2$}}}
\put(55.00,36.00){\makebox(0,0)[cc]{{\bf k$_1$}}}
\put(85.00,36.00){\makebox(0,0)[cc]{{\bf k$_2$}}}
\put(70.00,27.00){\makebox(0,0)[cc]{{\bf k$_3$}}}
\put(53.00,11.00){\makebox(0,0)[cc]{{\bf k$_4$}}}
\put(73.00,11.00){\makebox(0,0)[cc]{{\bf k$_5$}}}
\put(90.00,11.00){\makebox(0,0)[cc]{{\bf k$_6$}}}
\put(55.00,31.00){\line(4,1){4.00}}
\put(55.00,31.00){\line(4,-1){4.00}}
\put(78.00,31.00){\line(4,1){4.00}}
\put(78.00,31.00){\line(4,-1){4.00}}
\put(67.00,23.00){\line(4,1){4.00}}
\put(67.00,23.00){\line(4,-1){4.00}}
\put(52.00,15.00){\line(4,1){4.00}}
\put(52.00,15.00){\line(4,-1){4.00}}
\put(70.00,15.00){\line(4,1){4.00}}
\put(70.00,15.00){\line(4,-1){4.00}}
\put(85.00,15.00){\line(4,-1){4.00}}
\put(85.00,15.00){\line(4,1){4.00}}
\put(28.00,23.00){\line(2,1){4.00}}
\put(28.00,23.00){\line(2,-1){4.00}}
\put(108.00,23.00){\line(2,1){4.00}}
\put(108.00,23.00){\line(2,-1){4.00}}
\put(70.00,31.00){\circle*{2.00}}
\put(80.00,15.00){\circle*{2.00}}
\put(62.00,15.00){\circle*{2.00}}
\put(61.50,15.00){\line(0,-1){15.00}}
\put(62.50,15.00){\line(0,-1){15.00}}
\put(70.00,35.00){\makebox(0,0)[cc]{{\bf O$_\mu$}}}
\put(80.00,19.00){\makebox(0,0)[cc]{{\bf O$_\mu$}}}
\put(60.00,19.00){\makebox(0,0)[cc]{{\bf $\Gamma_M$}}}
\put(62.00,-2.80){\makebox(0,0)[cc]{{\large\bf M}}}
\end{picture}
\begin{center}
\vspace*{1cm}
\hspace*{-1cm}Diagram II$_a$:
\end{center}

{\bf
\vspace*{.5cm}
\begin{center}
\begin{tabular}{ll}
k$_1$ = k$^\prime$+p$_2$ \hspace*{2cm} & k$_2$ = k+p$_1$\\
k$_3$ =  - (k$^\prime$ + k$^{\prime\prime}$) + p$_3$/2 \hspace*{2cm} &
k$_4$ = k$^{\prime\prime}$-p$_3$/2\\
k$_5$ = k$^{\prime\prime}$+p$_3$/2 \hspace*{2cm} &
k$_6$ = k$^{\prime\prime}$+k$^\prime$-k-p$_3$/2\\
\end{tabular}
\end{center}
}

\newpage
\unitlength=1.00mm
\special{em:linewidth 0.4pt}
\linethickness{0.4pt}
\begin{picture}(127.00,116.00)
\put(70.00,23.00){\oval(48.00,16.00)[]}
\put(46.00,23.00){\circle*{5.00}}
\put(94.00,23.00){\circle*{5.00}}
\put(95.00,23.00){\line(1,0){25.00}}
\put(120.00,24.00){\line(-1,0){25.00}}
\put(95.00,22.00){\line(1,0){25.00}}
\put(45.00,23.00){\line(1,0){50.00}}
\put(45.00,22.00){\line(-1,0){25.00}}
\put(20.00,23.00){\line(1,0){25.00}}
\put(45.00,24.00){\line(-1,0){25.00}}
\put(81.00,02.00){\line(0,3){8.00}}
\put(81.00,02.00){\line(-1,3){1.50}}
\put(81.00,02.00){\line(1,3){1.50}}
\put(86.00,05.00){\makebox(0,0)[cc]{{\bf p$_3$}}}
\put(13.00,23.00){\makebox(0,0)[cc]{{\large\bf B$_{Q^\prime(q)}$}}}
\put(127.00,23.00){\makebox(0,0)[cc]{{\large\bf B$_{Q}$}}}
\put(110.00,29.00){\makebox(0,0)[cc]{{\bf p$_1$}}}
\put(30.00,29.00){\makebox(0,0)[cc]{{\bf p$_2$}}}
\put(55.00,36.00){\makebox(0,0)[cc]{{\bf k$_1$}}}
\put(85.00,36.00){\makebox(0,0)[cc]{{\bf k$_2$}}}
\put(70.00,27.00){\makebox(0,0)[cc]{{\bf k$_3$}}}
\put(53.00,11.00){\makebox(0,0)[cc]{{\bf k$_4$}}}
\put(71.00,11.00){\makebox(0,0)[cc]{{\bf k$_5$}}}
\put(90.00,11.00){\makebox(0,0)[cc]{{\bf k$_6$}}}
\put(55.00,31.00){\line(4,1){4.00}}
\put(55.00,31.00){\line(4,-1){4.00}}
\put(78.00,31.00){\line(4,1){4.00}}
\put(78.00,31.00){\line(4,-1){4.00}}
\put(67.00,23.00){\line(4,1){4.00}}
\put(67.00,23.00){\line(4,-1){4.00}}
\put(52.00,15.00){\line(4,1){4.00}}
\put(52.00,15.00){\line(4,-1){4.00}}
\put(68.00,15.00){\line(4,1){4.00}}
\put(68.00,15.00){\line(4,-1){4.00}}
\put(85.00,15.00){\line(4,-1){4.00}}
\put(85.00,15.00){\line(4,1){4.00}}
\put(28.00,23.00){\line(2,1){4.00}}
\put(28.00,23.00){\line(2,-1){4.00}}
\put(108.00,23.00){\line(2,1){4.00}}
\put(108.00,23.00){\line(2,-1){4.00}}
\put(70.00,31.00){\circle*{2.00}}
\put(78.00,15.00){\circle*{2.00}}
\put(62.00,15.00){\circle*{2.00}}
\put(77.50,15.00){\line(0,-1){15.00}}
\put(78.50,15.00){\line(0,-1){15.00}}
\put(70.00,35.00){\makebox(0,0)[cc]{{\bf O$_\mu$}}}
\put(62.00,19.00){\makebox(0,0)[cc]{{\bf O$_\mu$}}}
\put(76.00,19.00){\makebox(0,0)[cc]{{\bf $\Gamma_M$}}}
\put(78.00,-2.80){\makebox(0,0)[cc]{{\large\bf M}}}
\end{picture}
\begin{center}
\vspace*{1cm}
\hspace*{-1cm}Diagram II$_b$:
\end{center}

{\bf
\vspace*{.5cm}
\begin{center}
\begin{tabular}{l l}
k$_1$ = k$^\prime$+p$_2$ \hspace*{2cm} & k$_2$ = k+p$_1$\\
k$_3$ =  - (k + k$^{\prime\prime}$) - p$_3$/2 \hspace*{2cm} &
k$_4$ = k$^{\prime\prime}$-k$^\prime$+k+p$_3$/2\\
k$_5$ = k$^{\prime\prime}$-p$_3$/2 \hspace*{2cm} &
k$_6$ = k$^{\prime\prime}$+p$_3$/2\\
\end{tabular}
\end{center}
}

\newpage
\unitlength=1.00mm
\special{em:linewidth 0.4pt}
\linethickness{0.4pt}
\begin{picture}(127.00,116.00)
\put(70.00,23.00){\oval(48.00,16.00)[]}
\put(46.00,23.00){\circle*{5.00}}
\put(94.00,23.00){\circle*{5.00}}
\put(95.00,23.00){\line(1,0){25.00}}
\put(120.00,24.00){\line(-1,0){25.00}}
\put(95.00,22.00){\line(1,0){25.00}}
\put(45.00,23.00){\line(1,0){50.00}}
\put(45.00,22.00){\line(-1,0){25.00}}
\put(20.00,23.00){\line(1,0){25.00}}
\put(45.00,24.00){\line(-1,0){25.00}}
\put(75.00,02.00){\line(0,3){8.00}}
\put(75.00,02.00){\line(-1,3){1.50}}
\put(75.00,02.00){\line(1,3){1.50}}
\put(80.00,05.00){\makebox(0,0)[cc]{{\bf p$_3$}}}
\put(13.00,23.00){\makebox(0,0)[cc]{{\large\bf B$_{Q^\prime(q)}$}}}
\put(127.00,23.00){\makebox(0,0)[cc]{{\large\bf B$_{Q}$}}}
\put(110.00,29.00){\makebox(0,0)[cc]{{\bf p$_1$}}}
\put(30.00,29.00){\makebox(0,0)[cc]{{\bf p$_2$}}}
\put(55.00,36.00){\makebox(0,0)[cc]{{\bf k$_1$}}}
\put(85.00,36.00){\makebox(0,0)[cc]{{\bf k$_2$}}}
\put(55.00,27.00){\makebox(0,0)[cc]
{{\bf k$_3$}}}
\put(85.00,27.00){\makebox(0,0)[cc]
{{\bf k$_4$}}}
\put(55.00,11.00){\makebox(0,0)[cc]
{{\bf k$_5$}}}
\put(85.00,11.00){\makebox(0,0)[cc]
{{\bf k$_6$}}}
\put(55.00,31.00){\line(4,1){4.00}}
\put(55.00,31.00){\line(4,-1){4.00}}
\put(55.00,23.00){\line(4,1){4.00}}
\put(55.00,23.00){\line(4,-1){4.00}}
\put(55.00,15.00){\line(4,1){4.00}}
\put(55.00,15.00){\line(4,-1){4.00}}
\put(78.00,31.00){\line(4,1){4.00}}
\put(78.00,31.00){\line(4,-1){4.00}}
\put(78.00,23.00){\line(4,1){4.00}}
\put(78.00,23.00){\line(4,-1){4.00}}
\put(78.00,15.00){\line(4,1){4.00}}
\put(78.00,15.00){\line(4,-1){4.00}}
\put(28.00,23.00){\line(2,1){4.00}}
\put(28.00,23.00){\line(2,-1){4.00}}
\put(108.00,23.00){\line(2,1){4.00}}
\put(108.00,23.00){\line(2,-1){4.00}}
\put(70.00,31.00){\circle*{2.00}}
\put(70.00,23.00){\circle*{2.00}}
\put(70.00,15.00){\circle*{2.00}}
\put(70.50,15.00){\line(0,-1){15.00}}
\put(69.50,15.00){\line(0,-1){15.00}}
\put(70.00,35.00){\makebox(0,0)[cc]{{\bf O$_\mu$}}}
\put(70.00,26.00){\makebox(0,0)[cc]{{\bf O$_\mu$}}}
\put(67.00,19.00){\makebox(0,0)[cc]{{\bf $\Gamma_M$}}}
\put(70.00,-2.80){\makebox(0,0)[cc]{{\large\bf M}}}
\end{picture}
\begin{center}
\vspace*{.5cm}
\hspace*{-1cm}Diagram III:
\end{center}

{\bf
\vspace*{.5cm}
\begin{center}
\begin{tabular}{ll}
k$_1$ = k$^\prime$+p$_2$ \hspace*{2cm} & k$_2$ = k+p$_1$\\
k$_3$ = - (k$^\prime$ + k$^{\prime\prime}$)+p$_3$/2 \hspace*{2cm} &
k$_4$ = - (k+ k$^{\prime\prime}$)-p$_3$/2\\
k$_5$ = k$^{\prime\prime}$-p$_3$/2 \hspace*{2cm} &
k$_6$ = k$^{\prime\prime}$+p$_3$/2\\
\end{tabular}
\end{center}
}

\end{document}